\documentclass[twoside,journal]{IEEEtran}
\usepackage{amsmath}   
\usepackage{amssymb}
\usepackage{mathrsfs}

\usepackage{balance}   

\usepackage{cite}      

\usepackage{ifpdf}     

\usepackage{multicol}
\usepackage{makeidx}
\usepackage{color}
\usepackage{epsf}
\usepackage{subfigure}
\usepackage{wrapfig}
\usepackage{fancyhdr}
\usepackage[lowtilde]{url}

\ifCLASSINFOpdf
   \usepackage[pdftex]{graphicx}
   \graphicspath{{./fig/}}
   \DeclareGraphicsExtensions{{.pdf},{.jpeg},{.jpg}}
\else
   \usepackage[dvips]{graphicx}
   \graphicspath{{./fig/}}
   \DeclareGraphicsExtensions{{.eps}}
\fi

\fancypagestyle{firstpage}{
  \fancyhf{}
  \renewcommand{\headrulewidth}{0pt}
  \renewcommand{\footrulewidth}{0pt}
  \fancyhead[RO,LE]{\small\thepage}
  \fancyhead[LO]{\small IEEE TRANSACTIONS ON COMMUNICATIONS, ACCEPTED FOR PUBLICATION}

  \fancyfoot[C]{}
  }

\def\refname{\normalsize \centerline{\bf REFERENCES} }

\DeclareMathAlphabet{\mathpzc}{OT1}{pzc}{m}{it}
\providecommand{\norm}[1]{\lVert#1\rVert}

\newcommand{\be}{\begin{eqnarray}}
\newcommand{\ee}{\end{eqnarray}}

\definecolor{CommentRed}{rgb}{0.75,0,0.10}
\definecolor{CommentBlue}{rgb}{0.10,0,0.75} 
\newcommand{\comment}[1]{\normalfont{#1}}  
\newcommand{\commentB}[1]{\normalfont{#1}} 

\begin{document}

\title{Noncoherent LDPC-Coded Physical-Layer  \\ Network Coding 
using Multitone FSK\vspace{-0.15cm} }


\author{
Terry Ferrett, \emph{Student Member, IEEE},  and 
Matthew C. Valenti, \emph{Fellow, IEEE}\\
\vspace{-0.35cm}
\thanks{Manuscript received May 5, 2017; revised October 4, 2017 and January 9, 2018; accepted January 23, 2018. Date of publication Month X, 2018; date of current version January 30, 2018. Portions of this paper were presented at the IEEE International Conference on Communications, 2013 and 2015.  The associate editor coordinating the review of this paper and approving it for publication was M. Ardakani.
  
  T. Ferrett and M. C. Valenti are with the Lane Department of Computer Science and Electrical Engineering, West Virginia University, Morgantown, WV 26506 USA (email: terry.r.ferrett@ieee.org; valenti@ieee.org).
  
  Color versions of one or more of the figures in this paper are available online at http://ieeexplore.ieee.org.
  
  Digital Object Identifier X.}
}


\maketitle


\begin{abstract}
\textbf{  A noncoherent two-way relaying system is developed using physical-layer network coding for improved throughput over conventional relaying in a fading channel. Energy-efficient noncoherent operation is achieved using multitone frequency shift keying (FSK). A novel soft-output demodulator is developed for the relay, and corresponding achievable exchange rates are found for Rayleigh fading and AWGN channels. Bit-error rate performance approaching the achievable rate is realized using a capacity-approaching channel code and a receiver architecture that iterates between demodulation and channel decoding. Iterative decoding is performed feeding information back from the channel decoder to the demodulator. Additionally, error-rate performance is made to approach the achievable rate more closely by optimizing LDPC codes for this system. The energy efficiency improvement obtained by increasing the modulation order is more dramatic for the proposed physical-layer network coding scheme than it is for a conventional point-to-point system. Using optimized LDPC codes, the bit-error rate performance is improved by as much as 1.1 dB over a widely known standardized LDPC code, and comes to within 0.7 dB of the limit corresponding to the achievable rate. Throughout this work, performance for physical-layer network coding is compared to conventional network coding. When noncoherent FSK is used, physical-layer network coding enables higher achievable rates, and conventional network coding exhibits better energy efficiency at low rates. }
\end{abstract}


\maketitle




\thispagestyle{firstpage}

\pagestyle{fancy}
\fancyhf{}
\fancyhead{}
\fancyhead[RO,LE]{\small\thepage}
\fancyhead[LO]{\small IEEE TRANSACTIONS ON COMMUNICATIONS, ACCEPTED FOR PUBLICATION}
\fancyhead[RE]{\small FERRETT et al.: NONCOHERENT LDPC-CODED PHYSICAL-LAYER NETWORK CODING}
\renewcommand{\headrulewidth}{0pt}
\renewcommand{\footrulewidth}{0pt}


\section{Introduction}

\PARstart{S}{uppose}
\commentB{
 two \emph{terminals} need to exchange information wirelessly, but are out of radio range.
Suppose further that an additional terminal is in range of both terminals that need to exchange information.
The additional terminal can be used as a \emph{relay} to establish communication,
a topology known as the \emph{two-way relay channel}.
For ease of exposition, suppose multiple access is implemented by time division.
Using conventional techniques, four time slots are required to exchange information between the terminals: two for transmission
to the relay and two more for the relay to transmit to each terminal.
\emph{Network coding} \cite{ahlswede:2000} reduces the requirement to three or even two time slots per exchange.
The reduction to three time slots is achieved by the relay combining the signals received from the terminals and broadcasting the combination such that each terminal can detect the other's information\cite{hausl:2006}.
Reduction to two time slots is accomplished by allowing the terminals to transmit to the relay at the same time and in the same band, deliberately interfering, a technique termed \emph{physical-layer network coding} (PNC) \cite{zhang2:2006}.
}

\commentB{
Now suppose that coherent detection is difficult or impractical.
Fast-frequency hopping systems \cite{2011:torrieri_princ} and high-speed wireless receivers with significant Doppler such as trains are examples where coherent reception is challenging.
Performing coherent reception at the relay in the two-way relay channel is even more challenging than a conventional point-to-point channel since the network contains three \emph{oscillators} that must be synchronized, one at each terminal and one at the relay.
While the relay receiver could lock to the phase of one of the two terminal signals, the other will always be received with some (possibly time-varying) phase offset \cite{wu:2014}.
Relaxing the need for coherent reception using \emph{noncoherent} techniques is a fundamental problem for the PNC two-way relay channel.
}

\commentB{
It is well-known that \emph{frequency shift keying} (FSK) is an energy-efficient modulation that enables noncoherent reception.
When energy-per-bit is held constant, increasing FSK modulation order improves energy efficiency by increasing the distance between constellation points as a function of energy-per-bit.
Additionally, FSK exhibits a constant envelope, allowing the use of inexpensive nonlinear amplifiers, and can be implemented to have continuous phase (CPFSK) yielding a more compact spectrum.
Prior art has focused on developing binary FSK \cite{sorensen:2009, vtf:2011, zhang:2014, dang:2016} or coherent multitone (i.e., M-ary) FSK receivers \cite{yu:2016} for the PNC two-way relay channel. To our knowledge, no prior work (other than our related conference papers \cite{ferrett:2011,ferrett:2013,ferrett:2015}, which we discuss below) has considered noncoherent M-ary FSK, which is the focus of the present work. An alternative to noncoherent FSK is differential phase-shift keying (DPSK) \cite{cui:2009, zhu:2012}, however, DPSK is more sensitive to Doppler and frequency instability than noncoherent FSK \cite{2011:torrieri_princ}.
}

\commentB{
Several fundamental questions remain unanswered about the performance limits for systems that use noncoherent M-ary FSK to communicate over the PNC two-way relay channel.  In order to investigate these limits, we develop a \emph{soft-output noncoherent FSK demodulator} and determine the \emph{achievable rate} when using it for a variety of channel conditions.  To realize a system having performance that closely approaches the achievable rate, we utilize a \emph{capacity-approaching channel code} and develop a receiver that iterates between demodulation and channel decoding.  Given this architecture, another fundamental question is whether performance can be improved over off-the-shelf, standardized channel codes.  To address this question, we optimize channel codes for this architecture.
While PNC improves throughput over conventional three-step network coding, the exact throughput improvement is not known.  To determine the improvement, we  compare the achievable rate and channel-coded performance for both.
}



\commentB{
In general, there are several approaches to combining channel coding and PNC\cite{zhang:2008},
based on whether decoding is performed at the relay, terminals, or both.
In this work we consider the model where channel coding is performed at both the relay and terminals
over \emph{network-coded bits} using \emph{bit-interleaved coded modulation with iterative decoding} (BICM-ID) \cite{caire:1998} \cite{li:1997}.
We consider mapping the received symbols to network-coded bits at the demodulation step, which has been shown to discard information compared to other mapping strategies\cite{zhang:2009}, however, applying iterative decoding between the demodulator and decoder mitigates some performance
loss \cite{li_zhang:2013}.  Additionally, optimizing channel codes for particular channel types and modulations yields
performance benefits \cite{valenti_xiang:2012}.
Optimizing LDPC degree distributions for the two-way relay channel using density evolution improves performance over codes designed for point-to-point channels \cite{tanc:2013, huang:2013}.
In this work, we optimize the channel coding scheme for the uplink stage from the terminals to the relay
using \emph{extrinsic information transfer} (EXIT) charts \cite{brink:2004} to determine degree distributions yielding improved performance.
}

\commentB{
Our main contribution is developing a \emph{noncoherent modulation and channel coding system} for the faded two-way relay channel with PNC for improved throughput, incorporating unique features that are not present in previous approaches.
The primary distinguishing feature is removing the need for carrier phase synchronization while achieving capacity approaching performance using noncoherently-detected M-ary frequency shift keying (FSK), formulated for iterative soft-output channel decoding.  FSK energy efficiency improves as modulation order is increased, and satisfyingly, one of our key results is that the energy efficiency improvement when using high-order FSK rather than binary FSK is greater in a PNC system than in a single-terminal point-to-point system.  To determine the suitability of each protocol, we compare noncoherent PNC and conventional three-step network coding.  Notable conclusions include:
}

\begin{enumerate}
\item \commentB{Achievable rate results indicate that for PNC, quaternary modulation exhibits an energy
efficiency gain over binary modulation of up to $3$ dB in AWGN and $4$ dB in Rayleigh fading.}
These gains are greater than for a point-to-point system, where $2$ dB and $2.5$ dB are gained in AWGN
and Rayleigh fading, respectively \cite{valenti:2005}.
\item Optimized LDPC codes exhibit up to $1.1$ dB energy efficiency improvement over standard codes,
and approach the limit predicted by achievable rate analysis by between $0.3$ and $1$ dB.
Improvement is proportional to modulation order.
\item For each combination of modulation order and channel, there exists a rate above which
PNC is always more energy efficient than conventional three-step network coding, and below which the opposite is true.    
\end{enumerate}

The rest of this paper is organized as follows.
Section II develops the system model.
Section III formulates the PNC relay demodulator.
\commentB{Section IV determines the achievable rate for the two-way relay channel considering both PNC and three-step network coding.}
\commentB{Section V presents simulated bit error rate performance, 
the LDPC code degree distribution optimization technique for PNC and optimization results.}
Section VI provides concluding remarks.

\section{System Model}\label{sec:sysm}

\commentB{
We consider a two-step exchange for the two-way relay channel (TWRC) where the \commentB{terminals} transmit to the relay during the \emph{multiple-access} (MA) stage, and the relay then broadcasts to the \commentB{terminals} during the \emph{broadcast} (BC) stage.
A primary distinction between PNC schemes is whether the relay \emph{decodes-and-forwards}
or \emph{amplifies-and-forwards} the signal it receives during the MA stage.
We consider decode-and-forward, and to emphasize the relay decoding operation we refer to our PNC scheme as \emph{digital network coding} (DNC).
Additionally, the conventional three-step network coding scheme where the \commentB{terminals} transmit in separate times and bands 
is referred to as \emph{link-layer network coding} (LNC), as the network coding operation is performed above the physical layer.
}


\vspace{-4mm}
\subsection{\commentB{Multiple-Access Stage}}
\begin{figure}[t]
\centering
\vspace{2.5mm}
\includegraphics[width=\columnwidth]{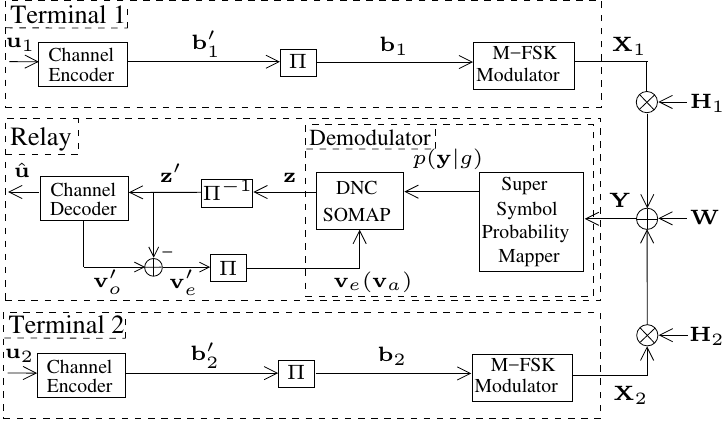}
\caption{System Model - Two-Way Relay Channel Digital Network-Coded (DNC) Multiple-Access stage}
\label{fig:sysm}
\vspace{-0.6cm}
\end{figure}

\commentB{
The system model for the DNC multiple-access stage is shown in Fig. \ref{fig:sysm}.
}
Two \commentB{\emph{terminals}} $\mathcal{N}_i,\ i \in \{1, 2\}$ each generate length-$K$ binary information sequences $\mathbf{u}_i = [ u_{0,i}, ..., u_{K-1,i}]$.
Each $\mathbf{u}_i$ is encoded by a binary LDPC code having rate $r$, yielding codewords $N = K/r$.
The codeword is passed through an interleaver, modeled as a permutation matrix $\mathbf{\Pi}$ having dimensionality $N \times N$, yielding $\mathbf{b}_{i} = \mathbf{b}'_{i} \mathbf{\Pi}$.
We assume a vector channel model where the vector dimensions correspond to matched filter outputs,
each representing a particular FSK frequency.
The number of bits per symbol is $\mu = \log_2 M$, where $M$ is the modulation order.
The codeword $\mathbf{b}_{i}$ at each \commentB{terminal} is divided into $L = N/\mu$ sets of bits $\mathbf{b}_{k,i}$, $0 \leq k \leq L-1$.
Each set of $\mu$ codeword bits is mapped to an $M$-ary symbol $q_{k,i} \in \mathcal{D}$ according to a natural mapping,
where $k$ denotes the symbol period, $i$ denotes the \commentB{terminal}, and $\mathcal{D}$ is the set of all symbols $0 \leq q_{k,i} < M-1$.
\comment{To ensure that the FSK tones are orthogonal for noncoherent detection, the frequency separation between each tone is $\Delta f = 1/T$, where $T$ is the symbol period \cite{proakis:2008}.}
The transmitted channel symbols are represented by the set of
column vectors $\mathbf{x}_{k,i}$.
Each $\mathbf{x}_{k,i}$ is length M, contains a 1 at vector position $q_{k,i}$ and 0 elsewhere.

\begin{figure}[t]
\centering
\vspace{2.5mm}
\includegraphics[width=\columnwidth]{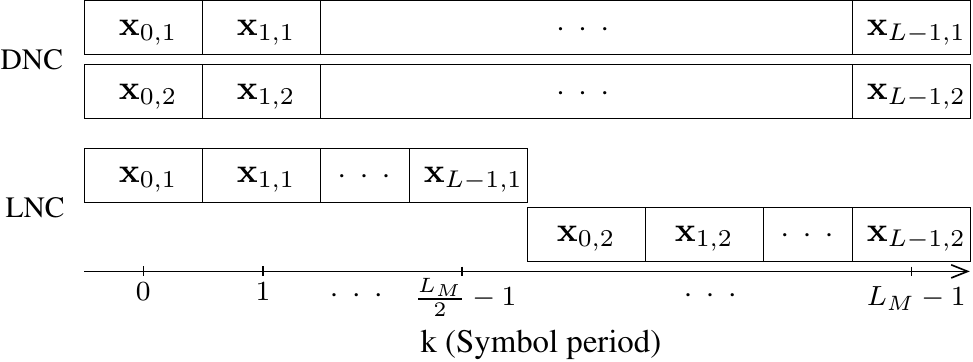}
\caption{Frame structure for digital and link-layer network coding (DNC and LNC) during the TWRC multiple-access (MA) stage.
The transmitted symbols are denoted by $\mathbf{x}_{a,b}$, where $a$ is the symbol period and $b \in \{1, 2\}$ denotes the \commentB{terminal}.  For DNC, each \commentB{terminal} transmits during the entire MA stage.  For LNC, the first \commentB{terminal} transmits during the first half of the MA stage, while the second \commentB{terminal} transmits during the second half.}
\label{fig:lnc_dnc_frames}
\vspace{-0.7cm}
\end{figure}

In order to fairly compare the error rate performance for the DNC and LNC protocols, the number of information bits $K$ transmitted to the relay by each \commentB{terminal} during the MA stage is assumed the same for both. Additionally, the duration in symbol periods allocated to both protocols is identical and denoted by $L_M$. 
Considering DNC, the \commentB{terminals} both transmit during the entire MA stage duration, thus, for the DNC case, $L = L_M$.
Considering LNC, each \commentB{terminal} is allocated half the MA stage duration, thus $L = L_M/2$ for LNC.
The relationship between the frame lengths for both protocols is shown in Fig. \ref{fig:lnc_dnc_frames}.

Define the \emph{MA rate} as the total number of network-coded information bits received at the relay during the MA stage
divided by the total number of bit periods $r_M = K/N_M$, where $N_M = \mu L_M$.
Since the \commentB{terminals} in the DNC case transmit during the entire MA stage, the codeword length
and code rate are $N=N_M$ and $r = r_M$ respectively.
The LNC \commentB{terminals} equally share the MA stage duration, and thus must use twice the rate as DNC to transmit the same number of information bits.
Thus, in LNC $N=N_M/2$ and $r = 2r_M$.
Performance is compared for DNC and LNC by assuming identical MA rates for both.

The modulated codeword transmitted by \commentB{terminal} $\mathcal{N}_i$ is represented by the matrix of symbols $\mathbf{X}_i = [\mathbf{x}_{0,i},...,\mathbf{x}_{L_M-1,i}]$ having dimensionality $M \times L_M$.
In the DNC case, each \commentB{terminal} transmits during the entire MA stage, thus, all $L_M$ columns of $\mathbf{X}_1$ and $\mathbf{X}_2$ contain symbols.
For LNC, \commentB{terminal} $\mathcal{N}_1$ transmits during the first half of the MA stage and $\mathcal{N}_2$ transmits during the second half, thus, $\mathbf{X}_1$ contains symbols in columns $0 \leq k \leq L_M/2-1$ and zeros elsewhere, while $\mathbf{X}_2$ contains symbols in columns $L_M/2 \leq k \leq L_M-1$ and zeros elsewhere.  The frame structures for DNC and LNC are shown in Fig. \ref{fig:lnc_dnc_frames}.


\vspace{-4mm}
\subsection{Channel Model}

The gain from \commentB{terminal} $\mathcal{N}_i$ to the relay during the $k^{th}$ signaling interval is $h_{k,i} = \alpha_{k,i} e^{j \theta_{k,i}}$, where $\alpha_{k,i}$ is Rayleigh distributed for the fading channel and constant for AWGN, and $\theta_{k,i}$ is the phase, which is uniformly distributed between $\left[0, 2\pi \right)$.
In fading, the gains are independent and identically distributed (i.i.d.) for each symbol period, and their distribution is specified such that the amplitudes have unit energy $E[\alpha^2_{k,i}] = 1$.
For AWGN, $\alpha_{k,i}=1$.


\comment{
A fundamental assumption for our model is that the amplitude corruption and phase shift induced by the channel
is constant for a symbol period.  This assumption requires symbol periods that are \emph{less than or equal to the coherence time of the channel} $T \leq T_c$, where $T$ and $T_c$ are the symbol period and coherence time respectively.
Equivalently, the symbol rate must be greater than the inverse of the coherence time $r_s > 1/T_c$, where $r_s = 1/T$.
Coherence time is proportional to the inverse of the Doppler spread $T_c \approx 1/f_m$.
When the relative velocity between a \commentB{terminal} and the relay is $v$, the Doppler spread is $f_m = f_c (v/c)$, where $c$ is the speed of light and $f_c$ is the carrier frequency.
As an example, consider carrier frequencies $f_c = 2.4$ GHz, and suppose a \commentB{terminal} travels at $60$ km/h with respect to the relay.  Then the symbol rate must be greater than $r_s = 133$ symbols/s.
Since both \commentB{terminals} transmit during the multiple access stage, and we have assumed that their symbol rates are identical,
the symbol period used by both must be less than the coherence time experienced by the faster \commentB{terminal}.
}


\commentB{Frame synchronization is a critical consideration in systems employing DNC.
One approach to achieving frame synchronization is by setting a timing advance, as done in LTE \cite{dahlman:2011}.
When a synchronization technique such as timing advance is not available, propagation delays can be
made insignificant by limiting the symbol rate.
Suppose that \commentB{terminals} 1 and 2 lie $d_1 = 300$ meters and $d_2 = 600$ meters from the relay, respectively.
The propagation delays from each \commentB{terminal} to the relay are $t_1 = d_1/c$ and $t_2=d_2/c$ respectively, where $c$ is the
speed of light.
To make the propagation delay insignificant, we must ensure that the difference is much less than half a symbol period, so $T >> 2|t_1 - t_2|$ is required, where $T$ is the symbol period.
Continuing the example, the symbol period must satisfy  $2\ \mu s >> 2|(300-600)/c|$, \emph{limiting the rate} to approximately $250$ kilosymbols/s.
An alternative approach is to delay transmission by the \commentB{terminal} closer to the relay, however, this approach requires accurate distance tracking.
}

At the relay, the frames transmitted by the \commentB{terminals} in the DNC case are received perfectly overlapped in time.
For LNC, without loss of generality, it is assumed that the relay begins receiving the frame transmitted by $\mathcal{N}_2$ immediately after reception of $\mathcal{N}_1$'s frame ends.
The received signal at the relay during the MA stage is
\begin{align} \label{eqn:dnc_fading}
\mathbf{Y} = \sqrt{\mathcal{E}_1} \mathbf{X}_1 \mathbf{H}_1 + \sqrt{\mathcal{E}_2}\mathbf{X}_2 \mathbf{H}_2 + \mathbf{W}
\end{align}
\noindent where $\mathcal{E}_1$ and $\mathcal{E}_2$ are the symbol energies transmitted by \commentB{terminals} $\mathcal{N}_1$ and $\mathcal{N}_2$ respectively, $\mathbf{H}_i$ is an $L_M \times L_M$ diagonal matrix of channel coefficients having value $h_{k,i}$ at matrix entry $(n,n)$ and $0$ elsewhere and $\mathbf{W}$ is a noise matrix having dimensions $M \times L_M$.
Each column of $\mathbf{Y}$ represents a \emph{channel observation} denoted by $\mathbf{y}_{k}$, where $k$ denotes the symbol period.
The $k^{th}$ column of $\mathbf{W}$, denoted as $\mathbf{w}_k$,
is composed of zero-mean circularly symmetric complex jointly Gaussian random variables having covariance matrix $N_0 \mathbf{I}_M$; i.e., $\mathbf{w}_k \sim \mathcal{N}_c(\mathbf{0}, N_0 \mathbf{I}_M)$.
$N_0$ is the one-sided noise spectral density, and $\mathbf{I}_M$ is the $M$-by-$M$ identity matrix.

\commentB{
  In a practical system, the carrier frequencies at the \commentB{terminals} and relay will not be perfectly synchronized due to oscillator offset.
  Synchronization is even more difficult in the DNC system than in a conventional point-to-point system since the network
  contains three oscillators that must be synchronized, one at each network node.
  While compensation techniques for mismatched carrier frequency offset have been investigated, such as adjusting the relay oscillator to the average offset for both \commentB{terminal} oscillators \cite{wu:2014}, in this work we assume that the effect of offset is negligible, and establish conditions to satisfy this assumption.
}

\commentB{
The conditions we assume for negligible frequency offset are as follows.
The bandpass frequency for the $k$-th FSK tone is $f_k = f_c + k\Delta f$, where $k$ is an integer $0 \leq k \leq M-1$.
  Suppose the carrier frequency at a \commentB{terminal} is $f_c$ and at the relay is $f_c' = f_c + f_o$,
  where $f_o$ is the oscillator offset.
  The relay correlator for tone $\ell$ will have reference frequency $f_\ell' = f_c' + \ell\Delta f$.
  The frequency separation between the $k$-th tone transmitted by the \commentB{terminal} and the $\ell$-th correlator reference
  frequency at the relay is $\Delta f' = f_k - f_\ell' = f_o + (k-\ell)\Delta f$.
  The correlation between two FSK tones when no oscillator offset is present is proportional to $\text{sinc}(2 T(k-\ell)\Delta f])$\cite{proakis:2008}.
  In the oscillator offset scenario described above, the correlation between tones is proportional to
  \begin{align}\label{eq:corr2}
      A= \text{sinc}(2 [f_o/r_s + (k-\ell)]). 
  \end{align}
  \noindent When no other compensation is applied, in order to make Eq. (\ref{eq:corr2}) go to zero when $k \neq \ell$, the symbol rate $r_s$ must be much greater than the frequency offset ($r_s \gg f_o$).
}
\vspace{-4mm}
\subsection{Relay Reception}\label{subsec:relay_recep}


In the LNC case, the relay demodulates and decodes the codewords $\mathbf{b}_1$ and $\mathbf{b}_2$ transmitted by each \commentB{terminal} separately using conventional point-to-point techniques, yielding estimates of $\mathbf{u}_1$ and $\mathbf{u}_2$ that are
then added modulo-2 to form and estimate $\hat{\mathbf{u}}$ of the network-coded message $\mathbf{u}$.
While it is possible to detect the network-coded bits in LNC using a single channel decoding by log-likelihood ratio (LLR) arithmetic \cite{vtf:2011}, the error rate performance observed for separate decoding is considerably better, thus, we only consider separate decoding in this work.


For the DNC case, the received signal during each symbol period is the sum of symbols
transmitted by the \commentB{terminals}. The network-coded combination of codeword bits transmitted by
the \commentB{terminals} is defined as
\begin{align}
\mathbf{b} = [\ b_0(\mathbf{x}_{k,1}) \oplus b_0(\mathbf{x}_{k,2})\ ...\ b_{\mu- 1}(\mathbf{x}_{k,1}) \oplus b_{\mu-1}(\mathbf{x}_{k,2}) \  ]
\end{align}
where $b_m(x_{k,i})$ denotes the $m^{th}$ bit mapped to the $k$-th symbol transmitted by \commentB{terminal} $\mathcal{N}_i$.
The DNC relay demodulator computes the likelihoods of each network-coded bit. Since the
LDPC code is a linear code, the modulo-2 sum of transmitted bits forms a codeword from the
codebooks used by the \commentB{terminals}, thus, the channel decoding operation yields a decision on the
network-coded message bits $\mathbf{u}$.

The DNC relay demodulator takes as input the matrix of received symbols and a priori
probability (APP) LLRs of the network-coded bits and computes a posteriori LLRs that are passed
to the channel decoder.
The probabilities of receiving the symbols comprising the frame $\mathbf{Y}$ are
computed.
Exact details of the probability calculation are given in Section \ref{sec:demod}.
The symbol probabilities and a priori LLRs of the network-coded bits $\mathbf{v}_a$ are passed to the DNC SOMAP, which computes the a posteriori LLRs $\mathbf{z}$ for each network-coded bit in the frame.
The a posteriori LLR is deinterleaved to produce $\mathbf{z}' = \mathbf{z} \mathbf{\Pi}^{-1}$ and passed to the decoder.
The decoder refines the estimate of $\mathbf{z}'$, producing a posteriori LLRs $\mathbf{v}_o'$.  
The decoder input is subtracted from the decoder output to produce extrinsic LLR $\mathbf{v}_e' = \mathbf{v}_o' - \mathbf{z}'$ which is interleaved to produce $\mathbf{v}_e = \mathbf{v}_e' \mathbf{\Pi}$ and returned to the DNC SOMAP.
The decoder output becomes the demodulator a priori input $\mathbf{v}_e = \mathbf{v}_a$.
After the specified number of decoding iterations has completed, the relay computes an estimate $\hat{\mathbf{u}}$ of the network-coded information bits $\mathbf{u}$.


The average symbol signal-to-noise ratio $\mathcal{E}_i/N_0$ transmitted by each \commentB{terminal} is assumed to be known at the demodulator.
The demodulator may operate under several cases of channel state information (CSI): the coherent case in which the gains are completely known (full CSI), the case in which only the fading amplitudes $\alpha_{k,i}$ are known (partial CSI), and the case in which no information about the gains is known other than the average SNR (no CSI).

\vspace{-4mm}
\subsection{Broadcast Stage}

During the BC stage, he relay encodes and modulates the estimated network-coded message bits $\hat{\mathbf{u}}$ and broadcasts to the \commentB{terminals}.
The signal traverses two independent channels, and the \commentB{terminals} receive independently corrupted versions of the network-coded bits.
The \commentB{terminals} demodulate and decode the signal received from the relay to form estimates of $\hat{\mathbf{u}}$, $\bar{\mathbf{u}}$ at $\mathcal{N}_1$ and $\tilde{\mathbf{u}}$ at $\mathcal{N}_2$.
Each \commentB{terminal} estimates the information bits transmitted by the other \commentB{terminal} by subtracting its own information sequence from the sequence detected from the symbol transmitted by the relay: $\hat{\mathbf{u}}_2 = \bar{\mathbf{u}} \oplus \mathbf{u}_1$ at $\mathcal{N}_1$ and 
$\hat{\mathbf{u}}_1 = \tilde{\mathbf{u}} \oplus \mathbf{u}_2$ at $\mathcal{N}_2$.
Since the links from the relay to the \commentB{terminals} are conventional point-to-point links with no interfering transmissions, specific details for the \commentB{terminal} receivers are omitted.

\vspace{-3mm}

\section{Digital Network-Coded Relay Demodulator}\label{sec:demod}
       
The goal of the DNC relay demodulator is to map the received signal containing the sum of symbols transmitted by the \commentB{terminals} to LLRs associated with the network coded bits.
The demodulator operates iteratively, using information fed back from the channel decoder to refine LLRs during each decoding iteration.
After a specified number of iterations has been reached, the decoder makes a hard decision on the network-coded bits.

The demodulator processes a frame of received signals $\mathbf{Y}$ one observation at a time.
Since the operation performed on each observation is the same, we may drop the dependence
on a particular signaling interval in the frame to simplify the notation. Denote a single received
channel observation as $\mathbf{y}$.
During the first demodulation and decoding iteration, the demodulator
computes the probability of receiving each possible combination of symbols transmitted by the
\commentB{terminals}: $p(\mathbf{y}|g)$, where $g$ is defined as the tuple
\begin{align}
g = (q_1, q_2)\ \ \ q_1,q_2 \in \mathcal{D} \ \ \ g \in \mathcal{G}
\end{align}
where $q_1$ and $q_2$ denote the indices of the symbols from \commentB{terminal} $\mathcal{N}_1$ and $\mathcal{N}_2$, respectively,
and $\mathcal{G} = \mathcal{D} \times \mathcal{D}$.
We refer to $g$ as a \emph{super-symbol} and the mapping from \commentB{terminal} symbols to super symbol as the \emph{super-symbol
probability mapping stage}.
Since the cardinality of $\mathcal{G}$ is $M^2$, the relay receiver computes $M^2$.
probabilities, versus a conventional point-to-point reception from a single \commentB{terminal} which only requires $M$ probabilities.

During each decoding iteration, the symbol probabilities and a priori LLRs $\mathbf{v}$ are transformed to the set of $\mu$ LLRs $\mathbf{z}$ associated with the network-coded bits mapped to super-symbols.
We refer to this operation as \emph{digital network-coded soft mapping} (DNC SOMAP) and the input-output
relationship is illustrated in Fig. \ref{fig:sysm}.
A general description of soft mapping for the point-to-point channel is given by \cite{benedetto:1998}.
The $k^{th}$ a priori input LLR to the demodulator representing the $k^{th}$ bit mapped to the super-symbol is referred to as the \emph{a priori demodulator information} and is related to the input distribution by
\begin{align}\label{eq:input_bit}
v_k = \log \frac{ P_I(b_k = 1) }{ P_I(b_k = 0) }, \ 0 \leq k \leq \mu-1
\end{align}
\noindent where $b_k$ is the $k^{th}$ network coded bit mapped to the super-symbol.
Considering BICM-ID, prior to the first decoding iteration, the bit probabilities are assumed to be equally likely, so
the a priori LLRs are set to zero.
The output LLR representing the $k^{th}$ bit mapped to the super-symbol is the \emph{a posteriori demodulator information} and is related to the output distribution by
\begin{align} \label{eq:somap_out_llr}
z_k = \log \frac{ P_O(b_k = 1) }{ P_O(b_k = 0) }, \ 0 \leq k \leq \mu-1 .
\end{align}
The DNC SOMAP output distribution is related to the input distributions by
\begin{align}\label{eq:somap_out_distribution_symbolic}
P_O(b_k=\ell) = \sum_{\begin{subarray}  (g: b_k(g) = \ell \end{subarray}}  p(\mathbf{y} |g) \prod_{\begin{subarray} jj=0 \\j \neq k \end{subarray}}^{\mu-1} P_I( b_j(g))
\end{align}
where the function $b_k(g)$ selects the $k^{th}$ network-coded bit associated with the super-symbol
$g$: $b_k(g) = b_k(q_1) \oplus b_k(q_2)$.
Substituting the specific values of the distribution Eq. (\ref{eq:input_bit}) into the expression for output Eq. (\ref{eq:somap_out_distribution_symbolic}),
\begin{align}\label{eq:somap_out_distribution_specific}
P_O( b_k = \ell) =  \sum_{\begin{subarray} (g: b_k(g) = \ell \end{subarray}}  p(\mathbf{y} |g)  \prod_{\begin{subarray} jj=0 \\j \neq k \end{subarray}}^{\mu-1}  \frac{ e^{ b_j(g) v_j} }{ 1+e^{v_j} }.
\end{align}
 The output LLR of the DNC-SOMAP may be found by combining Eq. (\ref{eq:somap_out_distribution_specific}) and Eq. (\ref{eq:somap_out_llr}):
 \begin{align}\label{eq:somap_out_llr_full}
 z_k = \sum_{b=0}^1  (-1)^{1-b} \log \left[ \displaystyle\sum_{\begin{subarray} (q: b_k(g) = b \end{subarray}} p(\mathbf{y} |g)  \prod_{\begin{subarray} jj=0 \\j \neq k \end{subarray}}^{\mu-1}   e^{ b_j(g)  v_j} \right]
 \end{align}
\noindent where the term $(1+e^{v_j})$ cancels in the ratio.


For numeric implementation, it is useful to simplify Eq. (\ref{eq:somap_out_llr_full}) using the \emph{max-star} operator
\begin{align}\label{eq:maxstar}
\underset{i}{\operatorname{max} \hspace{-0.5mm}*} \{ x_i \} = \log \left\{ \sum_i e^{ x_i } \right\}
\end{align}
\noindent where the binary max-star operator is  $\max*(x,y) = \max(x,y) + \log( 1 + e^{ -|x-y| } ) $ and
multiple arguments imply a recursive relationship; for example: $\max*(x,y,z) = \max*( x, \max*(y,z) )$.
Applying the max-star operator to Eq. (\ref{eq:somap_out_llr_full})
\begin{align} \label{eq:somap_out_llr_maxstar}
 z_k & =  \sum_{b=0}^1 (-1)^{1 - b}  \underset{\begin{subarray} (g: b_k(g) = b \end{subarray}}{\operatorname{max}  \hspace{-0.5mm} *} \left[ \log p(\mathbf{y} | g) + \sum_{\begin{subarray} jj=0 \\ j \neq k\end{subarray}}^{\mu-1} b_j(g) v_j\right].
\end{align}
\noindent The values taken by the pdf $p(\mathbf{y}|g)$ are dependent on the available channel state information.
Description of these pdfs is given in the following subsection.

\vspace{-4mm}
\subsection{Super-Symbol Probability Distributions}\label{subsec:distributions}
\vspace{-1mm}
\subsubsection{Full CSI}
When conditioned on the fading coefficients and transmitted signals, the output of the matched-filters is the sum of two $M$-dimensional complex Gaussian vectors, which is itself Gaussian.
Note that this receiver formulation is fully coherent.
Let $\mathbf{m}$ denote the mean of the received Gaussian vector when the symbols $q_1$ and $q_2$ are transmitted by the \commentB{terminals}.  There are $M^2$ such vectors, each having the form
\begin{eqnarray}
\mathbf{m} & = &  \sqrt{\mathcal{E}_1}h_1 \mathbf{x}_1 + \sqrt{\mathcal{E}_2}h_2 \mathbf{x}_2.
\end{eqnarray}
\comment{The super-symbol probability mapper computes $p(\mathbf{y}|g,\mathbf{h})$ for all values of $g$,
where $\mathbf{h} = [h_1 \ h_2]$.}
Applying the definition of the pdf of an M-dimensional complex-Gaussian vector, it is found that    
\comment{\begin{eqnarray}\label{Eqn_coherent}
p(\mathbf{y}|g,\mathbf{h}) = \left( \frac{1}{\pi N_0} \right)^M \exp \left\{ -\frac{1}{N_0} \norm{\mathbf{y} - \mathbf{m}}^2 \right\}. \label{eqn:coherent}
\end{eqnarray}}

\vspace{-4mm}
\subsubsection{Partial CSI}\label{subsubsec:nc_csi}
When the amplitudes of the fading coefficients are available at the receiver but the phases are not, the conditional pdf is found by marginalizing over the unknown phases of the received tones.
When the \commentB{terminals} transmit different symbols ($q_1 \neq q_2$), there will be two tones received, and therefore two phases to marginalize
\comment{\begin{eqnarray}
p(\mathbf{y}|g, \boldsymbol{\alpha}) 
\mspace{-10mu} & = & \mspace{-10mu}
\int_{0}^{2\pi} \int_{0}^{2\pi} p(\theta_{1}) p(\theta_{2}) p(\mathbf{y}|g,\mathbf{h}) d \theta_{1} d \theta_{2} \label{eqn:same_csi}
\end{eqnarray}
\noindent where $\boldsymbol{\alpha} = [\alpha_1 \ \alpha_2]$ is a vector whose elements are the magnitudes of the corresponding elements of $\mathbf{h}$ and it is assumed that the two received phases are independent.}
Substituting (\ref{eqn:coherent}) into (\ref{eqn:same_csi}), the conditional pdf becomes
\begin{eqnarray}\label{mupdf}\label{eqn:csi_different}
 p(\mathbf{y}|g,\boldsymbol{\alpha})  
& = & \frac{\beta}{4 \pi^2}  \prod_{i=1}^2 \int_0^{2 \pi} \exp \left\{ - \frac{|y_{q_i} - \sqrt{\mathcal{E}_i}\alpha_i e^{j \theta_{i} }|^2}{N_0} \right\} d \theta_{i}  \nonumber \\
& = & \beta \prod_{i=1}^2 \exp \left\{ - \frac{\mathcal{E}_i\alpha_{i}^2}{N_0} \right\}I_0 \left( \frac{2|y_{q_i}| \sqrt{\mathcal{E}_i}\alpha_i}{N_0}  \right) 
\end{eqnarray}
\noindent where the phases are assumed to be uniformly distributed, and
the factor
\begin{eqnarray}
\beta & = & \left( \frac{1}{\pi N_0} \right)^{M} \prod_{\begin{subarray}{c} k = 1 \\ k \neq \{q_1,q_2\}\end{subarray}  }^M \exp \left\{ - \frac{ |y_k|^2 }{ N_0 } \right\} \label{eqn:beta}
\end{eqnarray}
is common to all possible pairs of symbols and cancels in the LLR.

When the \commentB{terminals} transmit the same symbols ($q_1 = q_2$), the effects of channel fading may be modeled by a single term comprised of the sum of fading coefficients $h = \sqrt{\mathcal{E}_1}h_1 + \sqrt{\mathcal{E}_2}h_2 = \alpha e^{j \phi}$ with 
phase $\phi = \angle (\ \sqrt{\mathcal{E}_1}\alpha_1 e^{j \theta_1} + \sqrt{\mathcal{E}_2}\alpha_2 e^{j \theta_2})$ and
amplitude 
\begin{eqnarray}
\alpha & = & |\sqrt{\mathcal{E}_1}\alpha_1 \exp(j \theta_1) + \sqrt{\mathcal{E}_2}\alpha_2 \exp(j \theta_2)|. \label{eqn:alpha}
\end{eqnarray}
The conditional pdf is found by marginalizing over $\phi$,
\begin{eqnarray}
p(\mathbf{y}|g,\alpha) 
& = &
\int_{0}^{2\pi} p(\phi) p(\mathbf{y}|g,h) d \phi. \label{eqn:diff_csi}
\end{eqnarray}
Noting that $\phi$ is uniformly distributed, the conditional pdf becomes
\begin{eqnarray}\label{eqn:csi_same}
  p(\mathbf{y}|g,\alpha)  & = &  \frac{\beta}{2 \pi} \int_0^{2 \pi} \exp \left\{ - \frac{|y_\ell - \alpha e^{j \phi }|^2}{N_0} \right\} d \phi \nonumber \\
 & = & \beta \exp \left\{ - \frac{\alpha^2}{N_0} \right\} I_0 \left( \frac{2|y_\ell| \alpha}{N_0}  \right)
 \end{eqnarray}
\noindent where $\beta$ is given by (\ref{eqn:beta}) and $\ell=q_1=q_2$.
The value of the amplitude $\alpha$ depends on the values of $\alpha_1$ and $\alpha_2$ as well as the phases $\phi_1$ and $\phi_2$.
Since the values of the phases are not known, the receiver may approximate the unknown amplitude as $\alpha = \sqrt{\mathcal{E}_1\alpha_1^2 + \mathcal{E}_2\alpha_2^2}$ \cite{valenti:2009}.

\subsubsection{No CSI}\label{subsubsec:nc_nocsi}
When the relay only has knowledge of the average received energy, the conditional pdf of the received signal is marginalized over the fading amplitudes.  When the \commentB{terminals} transmit different symbols, there are two fading amplitudes to marginalize over, and the conditional pdf becomes
\begin{eqnarray}\label{alpha_marg_diff}
p(\mathbf{y}|g) = \int_{0}^{2\pi} \int_{0}^{2\pi} p(\alpha_{1}) p(\alpha_{2}) p(\mathbf{y}|g,\boldsymbol{\alpha}) d \alpha_{1} d \alpha_{2}
\end{eqnarray}
where it is assumed that in the Rayleigh fading case the $\alpha_i$ are independent, each with pdf
\begin{eqnarray}\label{rayleighalpha12}
p(\alpha_{i}) & = & 2 \alpha_{i} \exp (-\alpha^2_{i}).
\end{eqnarray}
Substituting (\ref{eqn:csi_different}) and (\ref{rayleighalpha12}) into (\ref{alpha_marg_diff}) yields 

\vspace{-5mm}
\begin{eqnarray}\label{eqn:nocsi_diff}
p(\mathbf{y}|g) =
\beta \prod_{i=1}^2 \frac{1}{\mathcal{E}_i} \left( \frac{1}{\mathcal{E}_i} + \frac{1}{N_0} \right)^{-1}
 \exp \left\{ \frac{|y_{q_i}|^2 \mathcal{E}_i}{N_0(N_0 + \mathcal{E}_i)} \right\}.
\end{eqnarray}

When the same tone is transmitted by both \commentB{terminals}, the marginalization is over the composite fading amplitude $\alpha$
\begin{eqnarray}\label{alpha_marg_same}
p(\mathbf{y}|g) & = & \int_{0}^{2\pi}  p(\alpha) p(\mathbf{y}|g,\alpha) d \alpha.
\end{eqnarray}
Recall that the tone is received over a fading channel with an equivalent complex-fading coefficient $h=\sqrt{\mathcal{E}_1}h_1+\sqrt{\mathcal{E}_2}h_2 = \alpha e^{j \phi}$.
The amplitude $\alpha$ is Rayleigh with pdf 
\begin{eqnarray}\label{rayleighalpha}
p(\alpha) 
& = & 
2 \alpha \exp ( - \alpha^2 ).
\end{eqnarray}
Substituting (\ref{eqn:csi_same}) and (\ref{rayleighalpha}) into (\ref{alpha_marg_same}) and evaluating yields 
\begin{eqnarray}\label{eqn:nocsi_same}
p(\mathbf{y}|g) = 
\beta \left( \frac{1}{\mathcal{E}_1 + \mathcal{E}_2}\right) \left( \frac{1}{\mathcal{E}_1 + \mathcal{E}_2} + \frac{1}{N_0} \right)^{-1} \times ... \nonumber \\\exp \left\{ \frac{|y_{\ell}|^2 (\mathcal{E}_1 + \mathcal{E}_2)}{N_0^2 + N_0 (\mathcal{E}_1 + \mathcal{E}_2)} \right\}.
\end{eqnarray}
\noindent where $\ell = q_1 = q_2$.

\section{\comment{Achievable Rate}}\label{sec:capacity}

In this section, the \comment{achievable rate} for the DNC and LNC protocols is analyzed and computed via simulation.
Specifically, expressions for \comment{achievable rate} suitable for Monte Carlo simulation are derived
and used to generate rate curves.
For the \comment{achievable rate} analysis in this section, we are concerned with a single symbol period, so dependence on symbol period $k$ is dropped to simplify the notation.

\vspace{-0.4cm}
\subsection{\comment{Achievable Exchange Rate Analysis}}\label{subsec:e2e_cap}

All communication is assumed to be half-duplex, thus,
the MA stage and BC stage occur in separate time sequences.
In the MA stage, the \commentB{terminals} transmit information to the relay, and the
relay detects the network-coded combination of information bits from the \commentB{terminals}.
In the BC stage, the relay broadcasts the network-coded bits to the \commentB{terminals}.
Each \commentB{terminal} detects the bits transmitted by the opposite \commentB{terminal} by performing
channel decoding on the network-coded bits and subtracting its own bits.



During the MA stage, the goal of the relay is to estimate the likelihood of the network-coded symbol mapped to the pair of symbols transmitted by the \commentB{terminals}.
The network-coded symbol is defined in terms of the bits mapped to the symbols transmitted by the \commentB{terminals} as
\begin{align}\label{eq:dnc_nc_sym}
q = d( \mathbf{b}_1 \oplus \mathbf{b}_2 )
\end{align}
\noindent where $d(\cdot)$ is a function that maps a bit sequence to its decimal representation, i.e. $d(10) = 2$, and $\mathbf{b}_1$ and $\mathbf{b}_2$ are the bits mapped to symbols $q_1$ and $q_2$ respectively as described in Section \ref{sec:sysm}.
That is, all pairs of symbols transmitted by the \commentB{terminals} having the same modulo-2 sum of bits map to one network-coded symbol.
The network-coded symbol $q$ takes values $0 \leq q < M-1$.
Note that the transformation given by Eq. (\ref{eq:dnc_nc_sym}) is isomorphic to addition over the Galois field GF(M).
The MA stage may then be modeled as a \emph{virtual single-input single-output channel} having input $q$ and channel output $\mathbf{y}_r$, and thus, the transition distribution is $p(\mathbf{y}_r|q)$ \cite{noori:2012}.
Assuming uniformly distributed binary information sequences at the \commentB{terminals}, the \comment{achievable rate} during the MA stage
is given by the \emph{conditional average mutual information} (AMI) $I(q;\mathbf{y}_r)$ \cite{caire:1998}.

During the BC stage, the relay broadcasts a network-coded symbol to the \commentB{terminals}, where the network-coded
symbol is mapped to a bit sequence and channel symbol in the same manner as the \commentB{terminal} symbols as described in Section \ref{sec:sysm}.
Assuming that the statistics of the channels between the relay and each \commentB{terminal} are identical, the broadcast stage
may be modeled as a conventional point-to-point channel having transition probability
$p(\mathbf{y}_e|q)$ and \comment{achievable rate} $I(q;\mathbf{y}_e)$, where $\mathbf{y}_e$ is the received signal at the \commentB{terminals}.
Since the capacity of FSK in the point-to-point channel is well known, we omit the corresponding derivation in this work.
A thorough treatment may be found in \cite{valenti:2005}.

The \comment{achievable exchange rate} is a function of the MA and BC \comment{achievable rates} and the fraction of time allocated to each stage.
Exchange may be modeled as a cascade of point-to-point links, thus,
from the \emph{max-flow min-cut theorem} \cite{zhang:2010}, the 
\comment{achievable exchange rate} is found to be 
\begin{align}\label{eq:e2e_cap}
R_E = \max_{t_m}\ \min\ \{\ t_m I(q;\mathbf{y}_r),\ (1 - t_m) I(q; \mathbf{y}_e)\ \}
\end{align}
\noindent where $t_m$ is the fraction of time assigned to the MA stage and $1-t_m$ is the fraction
of time assigned to the BC stage.
The \comment{achievable exchange rate} is maximized by allocating transmission time to each stage
such that time-scaled \comment{achievable rates} are equated
\begin{align}\label{eq:ma_time}
t_m I(q;\mathbf{y}_r) &= (1 - t_m) I(q; \mathbf{y}_e) \nonumber\\
t_m &= \frac{I(q; \mathbf{y}_e)}{I(q;\mathbf{y}_r) + I(q; \mathbf{y}_e)}.
\end{align}
Substituting (\ref{eq:ma_time}) into (\ref{eq:e2e_cap}) yields
\begin{align}\label{eq:e2e_cap_ami}
R_E= \frac{I(q; \mathbf{y}_r)I(q; \mathbf{y}_e)}{I(q;\mathbf{y}_r) + I(q; \mathbf{y}_e)}.
\end{align}

\begin{figure}[!t]
  \centering
  \includegraphics[width=\columnwidth]{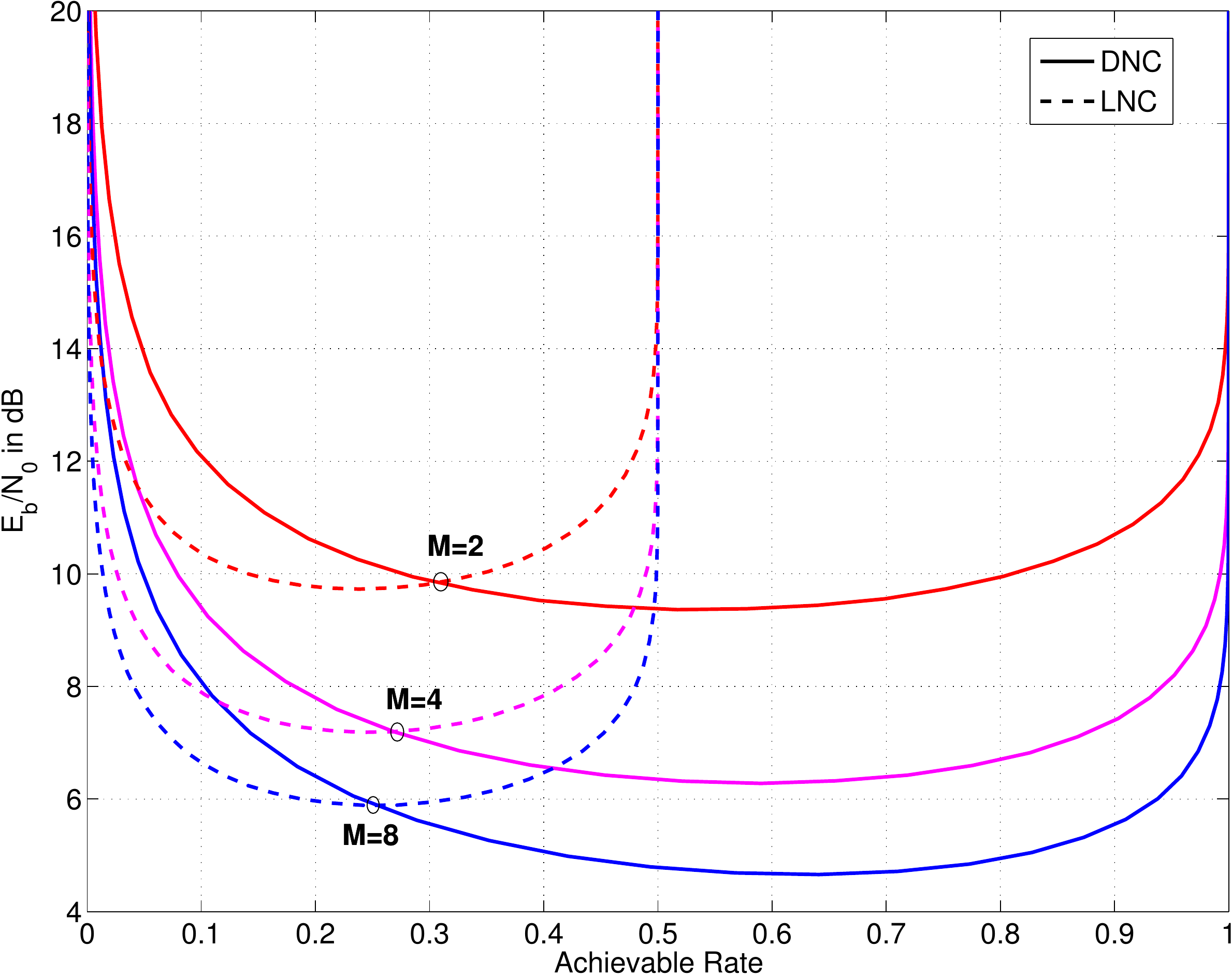}
  \caption{\comment{Achievable rate} for DNC
 and LNC multiple-access stages in AWGN with random phase noise.
 LNC takes rate values between 0 and 0.5, as each \commentB{terminal} requires separate time slots for transmission to the relay.
 Solid and dashed lines denote DNC and LNC respectively.
Modulation orders $M = \{ 2, 4, 8 \}$ are shown.
}
  \label{fig:cap_awgnrp}
  \vspace{-6mm}
\end{figure}

\vspace{-3mm}
\subsection{\comment{Achievable Rate} for the Multiple-Access Stage}

Considering DNC, the \commentB{terminals} transmit simultaneously to the relay during the MA stage, yielding the following received signal
at the relay for a single symbol period
\begin{align}
\mathbf{y}_r = \sqrt{\mathcal{E}_1}h_1 \mathbf{x}_1 + \sqrt{\mathcal{E}_2} h_2 \mathbf{x}_2 + \mathbf{w}
\end{align}
\noindent where $h_1$ and $h_2$ are complex channel gains and $\mathbf{w}$ is additive white Gaussian noise.
The transition distribution for this channel is $p(\mathbf{y}| g)$ as described in Section \ref{sec:demod}.
The achievable rate of the MA stage in DNC is given as
\begin{align}\label{eq:dnc_cm}
R_{D,M} = I(q; \mathbf{y}_r) = 1 - \frac{1}{\mu} E_{q, \mathbf{y}_r} 
\left[ \log_2 \frac{\sum\limits_{(q_1, q_2) \in g} p(\mathbf{y}_r| g) }
{ \sum\limits_{(q_1, q_2) \in g|_q}
 p(\mathbf{y}_r | g) }\right]
\end{align}
\noindent where $(q_1, q_2) \in g|_q$ denotes all combinations of symbols $q_1$ and $q_2$ such that $d(\mathbf{b}_1 \oplus \mathbf{b}_2) = q$, and in this paper rates are normalized by bits-per-symbol yielding units of information bit per code bit.

\begin{table}[t]
  \centering
  \caption{Most energy-efficient \comment{achievable rates} for the MA stage and corresponding SNR for DNC and LNC during the MA stage for AWGN and Rayleigh fading.  Each table entry takes form $(A:B)$, where $A$ is the \comment{achievable rate} and $B$ is the corresponding SNR ($E_b/N_0$ in $dB$).}
  \vspace{-4mm}
  \begin{tabular}{|l|l|lll|}
     \cline{1-5}
 Channel   & Protocol & $M=2$ & $M=4$ & $M=8$  \\
    \cline{1-5}
AWGN & DNC&$0.51:9.5$ & $0.6:6.2$ & $0.65:4.8$ \\
     \cline{2-5}      
    &  LNC&  $0.24:9.9$ & $0.25:7$& $0.25:5.9$ \\
     \cline{1-5}
    Rayleigh,  & DNC&$0.2:13$ & $0.25:9.7$ & $0.28:7.9$ \\
         \cline{2-5}      
    Partial CSI& LNC& $0.11:10.2$ &  $0.11:7.9$ &  $0.12:7.1$ \\ 
    \cline{1-5}
    Rayleigh, & DNC& $0.19:13.2$ & $0.22:10$ & $0.27:8.1$ \\
         \cline{2-5}      
    No CSI&  LNC& $0.14:11$  &  $0.14:8.3$  &  $0.14:7.5$\\
    \hline
  \end{tabular}
  \label{tbl:cap_summary}
  \vspace{-5mm}
\end{table}

In the LNC MA stage, the \commentB{terminals} transmit in separate time slots to the relay, 
yielding the pair of received symbols
\begin{align}\label{eq:lnc_rsc}
\mathbf{y}_{1} = \sqrt{\mathcal{E}_1}h_1 \mathbf{x}_1 + \mathbf{w}_1 \ \ \ \ \ \ \ \mathbf{y}_{2} = \sqrt{\mathcal{E}_2}h_2 \mathbf{x}_2 + \mathbf{w}_2
\end{align}
where $\mathbf{y}_1$ and $\mathbf{y}_2$ are the received signals from \commentB{terminals} $\mathcal{N}_1$ and $\mathcal{N}_2$
respectively.
The received signals in Eq. (\ref{eq:lnc_rsc}) may be modeled as two separate point-to-point channels during each time slot having transition distribution $p(\mathbf{y}_k|q_k),\ k \in \{1,2\}$.
We assume that each \commentB{terminal} is assigned one-half of the MA stage transmission time: $t_m/2$.
Thus, the MA \comment{achievable rate} for LNC may be modeled as a time-division multiple-access (TDMA) system where \comment{achievable rate} is one-half
that of a conventional point-to-point system \cite{cover:2006}.
The \comment{achievable rate} for the LNC multiple-access phase is
\begin{align}\label{eq:lnc_cm}
R_{L,M} =  \frac{1}{2}I(q_k; \mathbf{y}_k) = \frac{1}{2} - \frac{1}{2\mu} E_{q_k, \mathbf{y}_k} 
\left[ \log_2 \frac{\sum\limits_{q'_k \in \mathcal{D} } p(\mathbf{y}_k| q'_k) }
{ p(\mathbf{y}_k| q_k ) }\right]
\end{align}
\noindent where $I(q_k; \mathbf{y}_k)$ is the AMI for the point-to-point channel between \commentB{terminal} $\mathcal{N}_k$ and the relay, and the factor $\frac{1}{2}$ accounts for the TDMA characteristic of LNC.

\vspace{-3mm}
\subsection{Achievable Rate Results}


The \comment{achievable rate} for the MA stage is computed by simulation as follows.
A range of SNR values is specified, expressed as the ratio of symbol energy
to noise power $\mathcal{E}_s/N_0$, and $\mathcal{E}_s = \mathcal{E}_1 = \mathcal{E}_2$.
SNR is expressed in terms of energy per bit as
$\mathcal{E}_b/N_0 = \frac{\mathcal{E}_s/N_0}{C \log_2{M} }$, where $R$ denotes the \comment{achievable rate}.
Each \commentB{terminal} generates a binary information sequence and maps
it to $M$-ary FSK symbols as described in Section \ref{sec:sysm}.
The channel effects on the symbols transmitted from the \commentB{terminals} to the relay are simulated according
to Eq. (\ref{eqn:dnc_fading}).
The transmitted symbol energies are $\mathcal{E}_1 = \mathcal{E}_2 = 1$.

The achievable rate is computed using the received symbol frame and network-coded bit values. For DNC, the achievable rate is computed by substituting Eq. (\ref{eq:maxstar}) into Eq. (\ref{eq:dnc_cm}) resulting in
\begin{align}\label{eq:dnc_cm_maxstar}
R_{D,M} =1- \gamma  E_{q, \mathbf{y}}\left[ 
 \underset{\begin{subarray} ((q_1,q_2) \in g \end{subarray}}{\operatorname{max} \mspace{-2mu} *} \mspace{-5mu} \log p(\mathbf{y}| g)
 \mspace{-1mu} - \mspace{-10mu} \underset{\begin{subarray} ((q_1,q_2) \in g|_q \end{subarray}}{\operatorname{max}
\mspace{-2mu} *} \mspace{-10mu} \log p(\mathbf{y} | g) \right]
\end{align}
\noindent where $\gamma = \log_2(e)/\mu$. For LNC, the achievable rate is computed by substituting Eq. (\ref{eq:maxstar}) in Eq. (\ref{eq:lnc_cm}) resulting in
\begin{align}\label{eq:lnc_cm_maxstar}
R_{L,M} = \frac{1}{2} - \frac{\gamma}{2} E_{q_k, \mathbf{y}_{k}}\left[ 
 \underset{\begin{subarray} (q'_k \in \mathcal{D} \end{subarray}}{\operatorname{max}  \hspace{-0.5mm} *} \log p(\mathbf{y}_k| q'_k) - 
 \log p(\mathbf{y}_k | q_k) \right].
\end{align}
\noindent The expectations may be evaluated using Monte Carlo simulation and several hundred thousand trials.



The achievable rate in AWGN is shown in Fig. \ref{fig:cap_awgnrp}.
A summary of the \comment{achievable rates} for the MA stage which minimize the required SNR for AWGN are shown in Table \ref{tbl:cap_summary}.
The \comment{achievable rate} of DNC takes values between $0$ and $1$ while the \comment{achievable rate} of LNC takes values between
$0$ and $0.5$, demonstrating that DNC enables higher rates than possible for LNC.
At rates less than $0.5$, distinct regions exist where either DNC or LNC is more
energy efficient.
At modulation orders two, four and eight, LNC is more energy efficient at rates less than
approximately $0.3$, $0.27$ and $0.25$ respectively, while DNC is more efficient at rates higher than these values.
In general, the range of rates where DNC energy efficiency outperforms LNC increases with modulation order.
The performance gain between $M=2$ and $M=4$ is greater than between $M=4$ and $M=8$.
Generally, diminishing gains are observed as modulation order increases.

\begin{figure}[!t]
  \centering
  \includegraphics[width=\columnwidth]{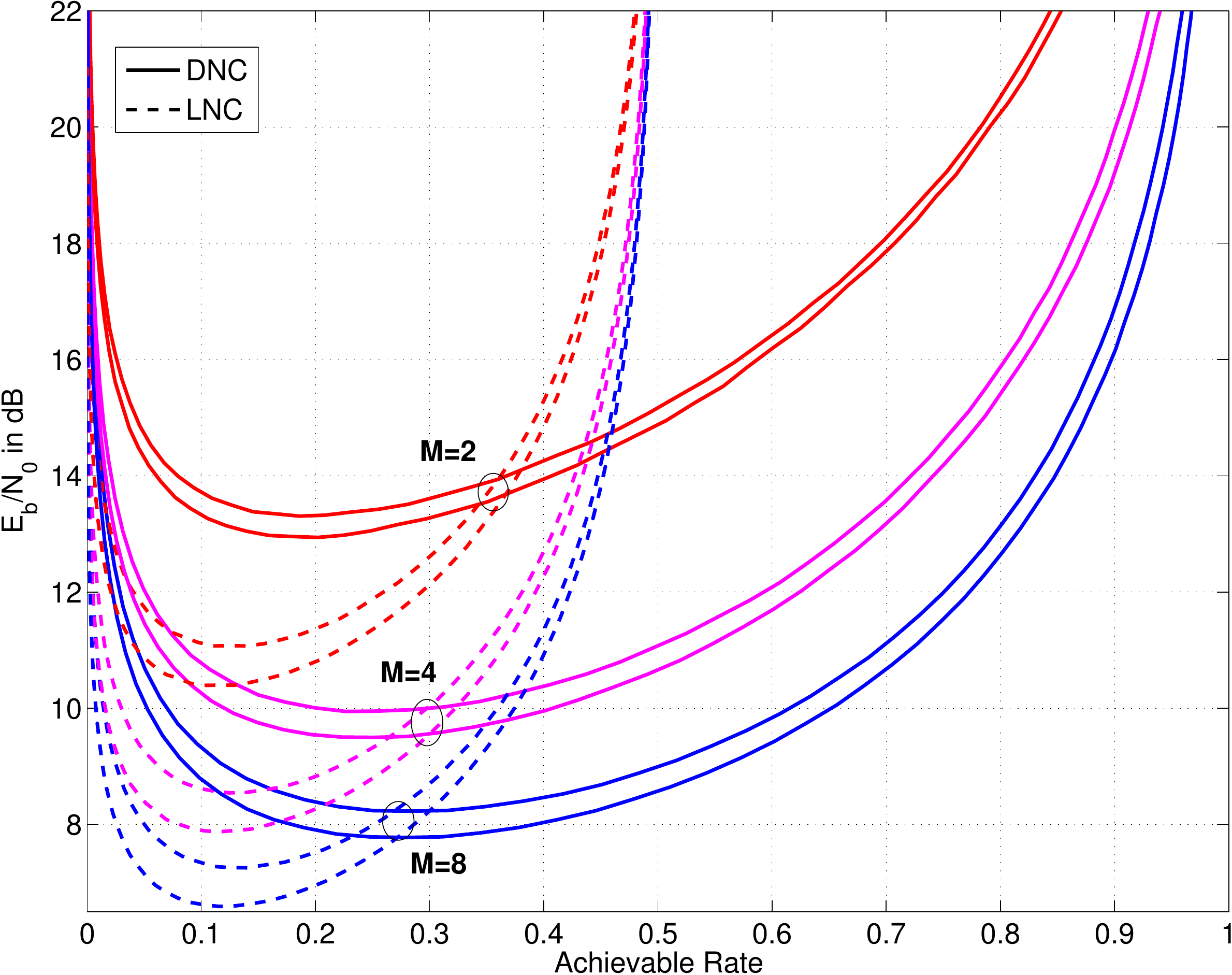}
  \caption{\comment{Achievable rate} for the digital network-coded (DNC)
 and link-layer network-coded (LNC) MA stage in Rayleigh fading.
 Modulation orders $M=\{2,4,8\}$ are considered.
 For DNC and LNC at every modulation order, within each pair of curves,
the upper and lower curves depict rate for partial and no CSI at the relay, respectively.
}
  \label{fig:cap_csi1csi2cm}
  \vspace{-5mm}
\end{figure}

\comment{Achievable rate} in Rayleigh fading with and without CSI at the relay is shown in Fig. \ref{fig:cap_csi1csi2cm}.
The maximum performance improvement of CSI over no CSI is approximately $1$ and $0.5$ dB for LNC and DNC respectively,
indicating that CSI is more beneficial for LNC.
Consider performance at rates less than $0.5$.
LNC exhibits better energy efficiency than DNC at approximate rates $0.35$, $0.3$, and $0.26$ when the relay has no CSI and rates $0.36$, $0.3$ and $0.27$ when the relay has partial CSI, at modulation orders two, four and eight, respectively.
LNC outperforms DNC over a wider range of rates in fading than in AWGN, demonstrating that fading degrades the \comment{achievable rate} of DNC more severely than LNC.
The MA rates which minimize the required SNR for Rayleigh fading are shown in Table \ref{tbl:cap_summary}.

\begin{figure}[!t]
  \centering
  \vspace{0.7mm}
  \includegraphics[width=\columnwidth]{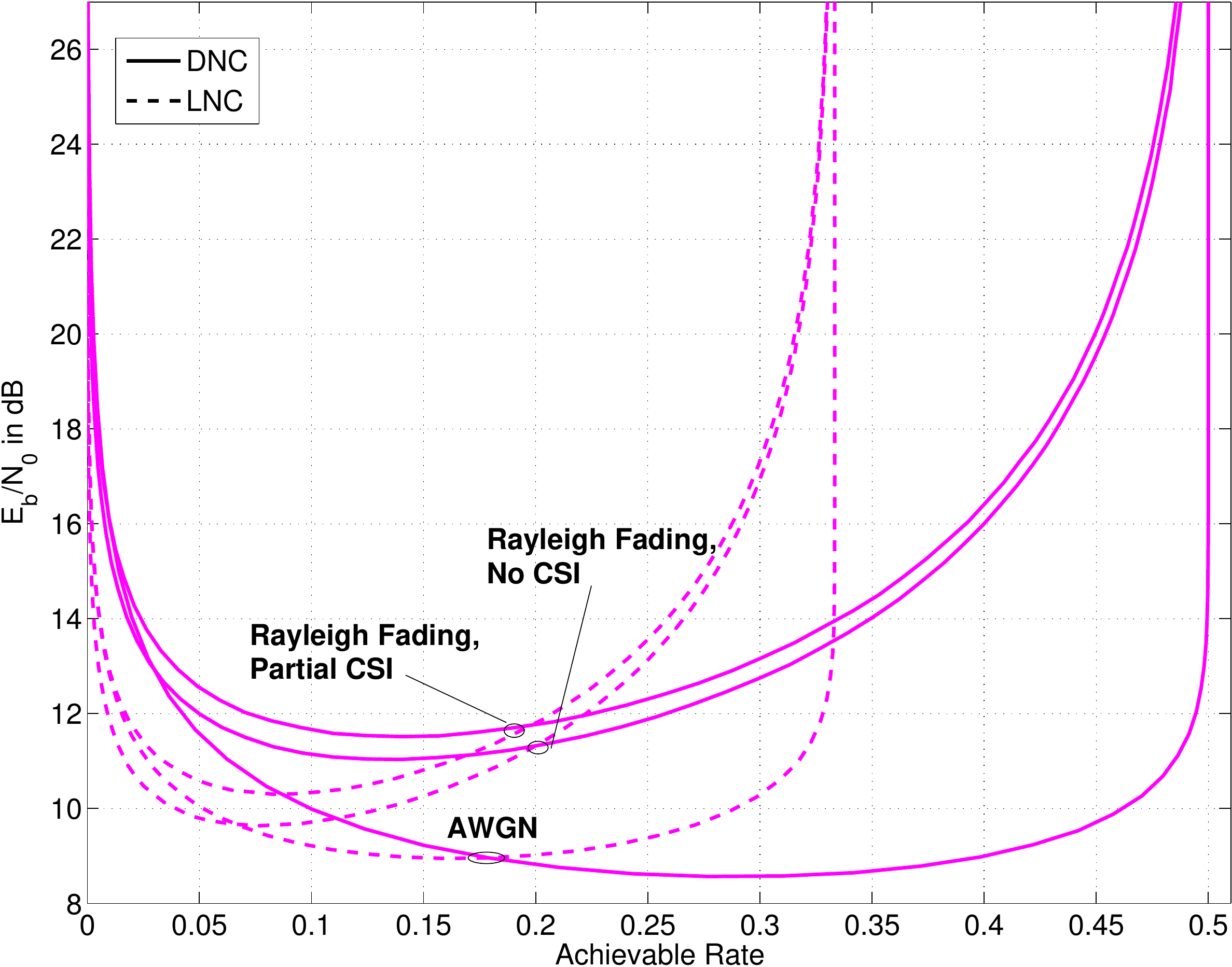}
  \caption{\comment{Achievable exchange rate} in AWGN and Rayleigh fading with no CSI and partial CSI for digital and link-layer network coding (DNC and LNC).  Results are shown for modulation order $M=4$.}
  \label{fig:cap_e2e}
  \vspace{-5mm}
\end{figure}

\comment{Achievable exchange rate} is shown in Fig. \ref{fig:cap_e2e}.
The rate is shown assuming partial and no CSI at the relay and for modulation order $M=4$.
The maximum rates for DNC and LNC are $1/2$ and $1/3$ respectively, which is consistent with DNC requiring two time slots to exchange information and LNC requiring three.
In AWGN, the energy efficiency of DNC outperforms LNC at rates greater than approximately $0.17$.
In fading, DNC outperforms LNC at rates greater than $0.18$ and $0.2$.
The maximum energy efficiency difference between partial and no CSI is about $0.5$ dB for LNC and about $0.25$ dB for DNC.

\section{LDPC Coded Performance and Optimization}\label{sec:ldpcopt}
\vspace{-1mm}

This section presents LDPC-coded error rate performance and optimization for the TWRC
multiple-access stage when combined with the DNC relay demodulator
described in Section \ref{sec:demod}.
Error rates are computed via Monte Carlo simulation.
Performance is investigated using the LDPC code defined by the DVB-S2 standard, and
the results are used as a baseline for optimization.
The optimization technique is based on matching the extrinsic information transfer (EXIT) characteristics 
of the demodulator and LDPC decoder.
Performance of the LNC protocol is simulated and compared to DNC.

\vspace{-4mm}
\subsection{Bit Error Rate Simulation Procedure}\label{subsec:ber_sim}

Throughout this section, the following procedure is applied to simulate LDPC-coded bit-error rate (BER) performance
during the MA stage.
A range of SNR values is specified, expressed as the ratio of bit energy
to noise power $\mathcal{E}_b/N_0$.
Each \commentB{terminal} generates a binary information sequence, performs LDPC encoding using the appropriate
parity check matrix to produce a codeword, and maps the codeword to $M$-ary FSK symbols as described in Section \ref{sec:sysm}.
The channel effects on the symbols transmitted from the \commentB{terminals} to the relay are simulated according
to Eq. (\ref{eqn:dnc_fading}).
The energy transmitted by each \commentB{terminal} is $\mathcal{E}_1 =\mathcal{E}_2  = 1$.

BER performance for standard codes is computed using parity check matrices defined by the DVB-S2 standard \cite{dvbs2:2013}.
In order to fairly compare performance between DNC and LNC during the MA stage, the number of information bits sent to the relay
by each \commentB{terminal} and the total number of symbol periods is assumed the same for both protocols.
The MA rates considered are $r_M = \{1/3,\ 2/5,\ 3/5 \}$, thus, $r = \{1/3,\ 2/5,\ 3/5 \}$ in the DNC case and $r = \{2/3,\ 4/5 \}$ in LNC.
The channel code lengths for DNC and LNC are $N=16200$ and $N=8100$ respectively.
The DVB-S2 standard does not specify codes having the length $N=8100$ as considered for LNC, thus, parity check matrices are generated having degree distributions identical to DVB-S2 codes with length and rates $16200$, $2/3$, and $4/5$ respectively.
The procedure for generating random parity check matrices is in subsection \ref{subsec:opt_vnd}.

Considering DNC the relay demodulates the received symbols using Eq. (\ref{eq:somap_out_llr_maxstar}) with
$p(\mathbf{y}|q)$ corresponding to the desired relay CSI as given in subsection \ref{subsec:distributions}.
BICM-ID decoding is performed as described in subsection \ref{subsec:relay_recep}.
The number of decoding iterations is 100 for all simulations, as further iterations do not significantly
improve decoding performance.
Considering LNC, demodulation for each subframe is performed by conventional point-to-point techniques as described in
\cite{valenti:2005}.
Several hundred thousand simulation trials are performed, and the number of network-coded bits in error is counted and used to compute the BER.

\vspace{-4mm}
\subsection{Channel-Coded Performance using Standard Codes}

The error rate is simulated as described in subsection \ref{subsec:ber_sim}.
The channel code is defined by the DVB-S2 standard \cite{dvbs2:2013}.
In fading, decoding is performed with and and without channel state information at the relay, as described in
the demodulator formulations given in subsections \ref{subsubsec:nc_csi} and \ref{subsubsec:nc_nocsi}.
LNC is simulated for comparison.

\begin{figure}[!t]
  \vspace{0.7mm}
  \centering
  \includegraphics[width=\columnwidth]{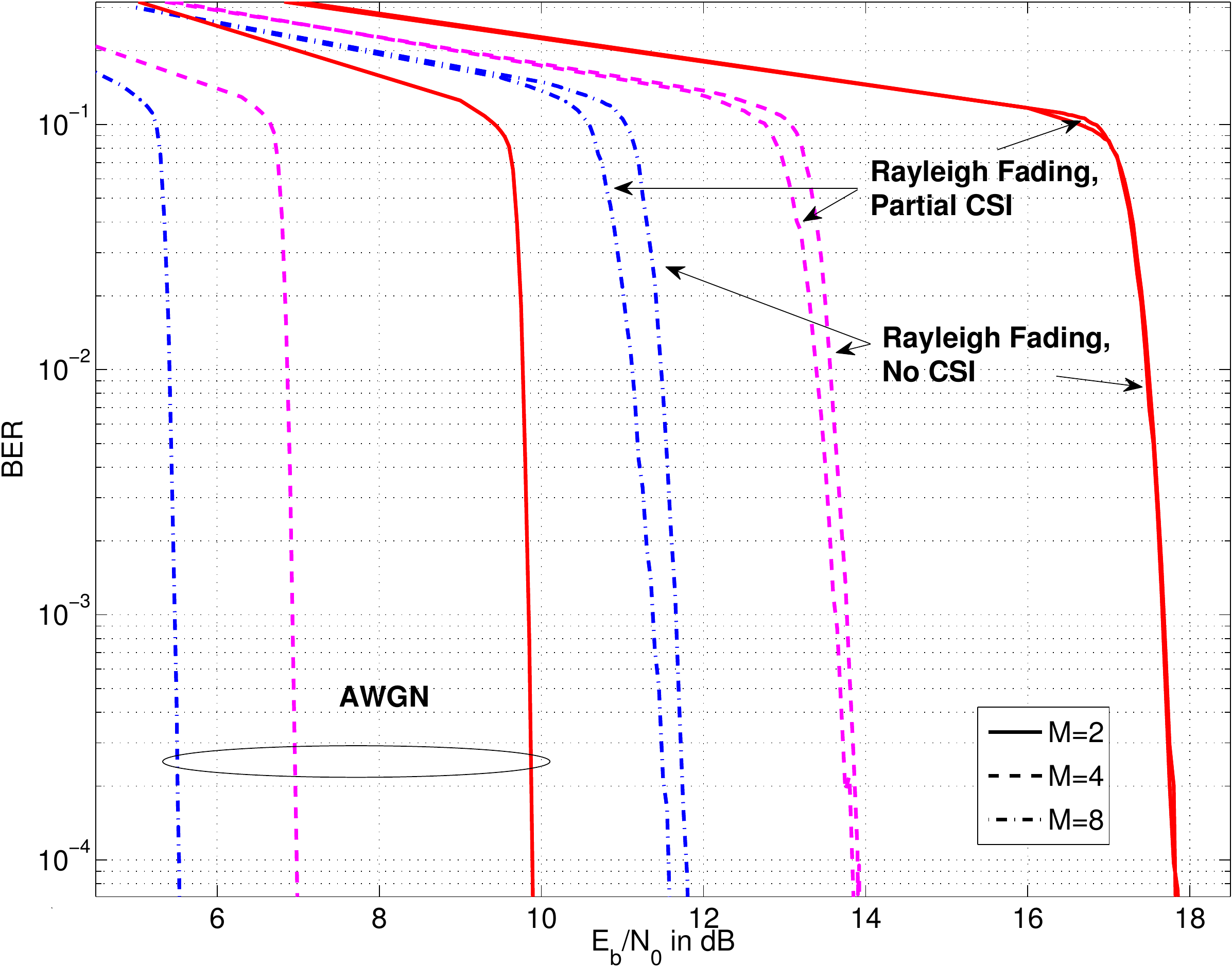}
  \caption{LDPC-coded BER performance at the relay for digital network coding (DNC) in AWGN and Rayleigh
fading channels using a DVB-S2 LDPC code.  The code length and rate are $N=16200$ bits and $r = 3/5$ respectively.  Results are shown for modulation orders $M=\{2,4,8\}$. In fading, performance with partial and no channel state information at the relay is shown.}
  \label{fig:dnc_r35}
  \vspace{-4mm}
\end{figure}

Error-rate performance for DNC at the relay using the DVB-S2 LDPC code having rate $r = 3/5$ is shown in Fig. \ref{fig:dnc_r35}.
This figure illustrates performance considering all channels and CSI cases and three modulation orders.
At modulation order $M=2$, the performance in fading is nearly identical regardless of the CSI available at the relay.
The difference in performance between partial and no CSI is about $0.1$ and $0.2$ dB at $M=4$ and $M=8$, respectively.
In fading, increasing modulation order from $2$ to $4$ and $4$ to $8$ improves energy efficiency by approximately $4$
and $2$ dB, respectively.
Similar behavior is observed in AWGN but with smaller performance differences between modulation orders.
A $3$ dB improvement when increasing modulation order from $2$ to $4$ and about $1$ dB of improvement when increasing from modulation order $4$ to $8$.
At an error rate of $10^{-4}$, in AWGN, the difference between BER performance and \comment{achievable rate} is about $0.5$, $0.6$ and $0.7$ dB for modulation orders $M=\{2,4,8\}$ respectively.
In fading, the difference between BER performance is about $1.5$, $2$ and $2$ dB respectively.

\begin{figure}[!t]
  \centering
  \includegraphics[width=\columnwidth]{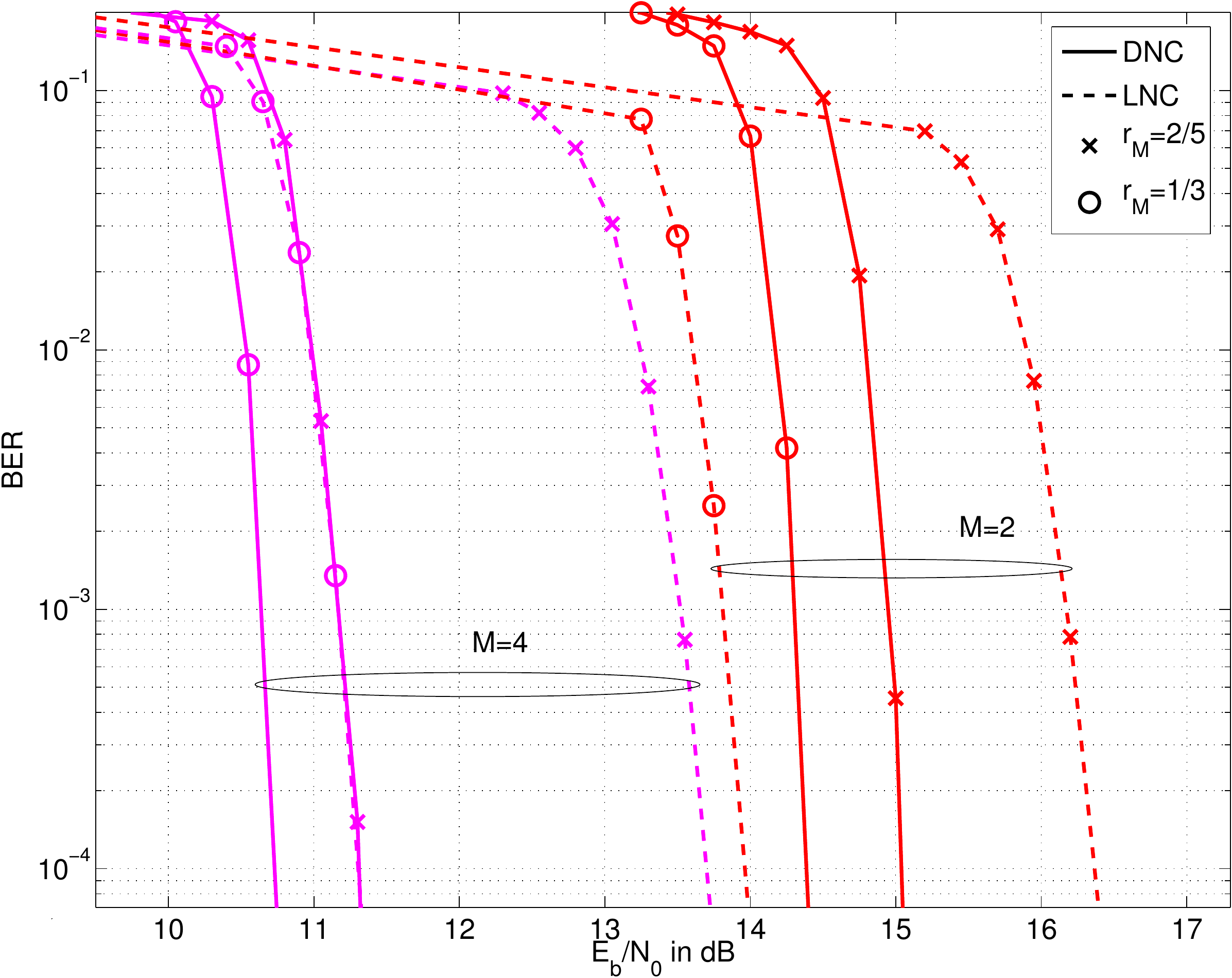}
  \caption{LDPC-coded BER performance at the relay for digital and link-layer network coding (DNC and LNC) in
Rayleigh fading at channel code rates $r_M=\{2/5,1/3\}$.  The relay possesses partial CSI as fading amplitudes.
The DNC and LNC frame lengths and rates are $N=16200$ and $N=8100$ bits, respectively. Results for modulation orders orders $M=\{2,4\}$ are shown.}
  \label{fig:dnclnc_twor}
  \vspace{-4mm}
\end{figure}

The bit-error rate performance of DNC and LNC at two different code rates is shown in Fig. \ref{fig:dnclnc_twor}.
All simulations consider Rayleigh fading with partial CSI at the relay demodulator.
At modulation order $M=2$ and MA rate $r_M=2/5$, DNC outperforms LNC by approximately $1.5$ dB, however at $r_M=1/3$ LNC outperforms DNC by about $0.5$ dB.
Considering modulation order $M=4$, DNC outperforms LNC by approximately $4$ dB at $r_M = 2/5$ and by roughly $0.5$ dB at $r_M=1/3$.
Generally, increasing modulation order increases the performance difference between DNC and LNC favorably for DNC.
At rate $r_M=2/5$, the difference between BER performance at $10^{-4}$ and \comment{achievable rate} is about $1.1$ dB for DNC and $1.5$ dB for LNC, on average.
At rate $r_M=1/3$, the difference is about $1$ dB and $0.7$ dB for DNC and LNC respectively.



\vspace{-4mm}
\subsection{LDPC Code Optimization}\label{subsec:opt_vnd}

An LDPC code may be fully described by a sparse binary parity check matrix $\mathbf{H}$.
The dimensionality of $\mathbf{H}$ is $N-K$ rows by $N$ columns.
Consider the \emph{Tanner graph} representation where the graph nodes are partitioned into two sets: \emph{variable nodes} and \emph{check nodes}.
The graph contains $N$ variable nodes and $N-K$ check nodes, one for each column and row of $\mathbf{H}$ respectively.
An \emph{edge} connecting the $n$-th variable node to the $k$-th check node corresponds to a $1$ in the parity check matrix located at row and column $(k,n)$.
The \emph{degree} of a variable or check node is the number of $1$'s in column $n$ of $\mathbf{H}$,
and the degree of check node $k$ is the number of $1$'s in row $k$ of $\mathbf{H}$.

LDPC variable and check nodes may be modeled as \emph{a posteriori probability} decoders which convert 
\emph{a priori} input LLRs to \emph{extrinsic} output LLRs  \cite{brink:2004}.
The transfer characteristics of the variable and check node decoders may be characterized by
measuring the mutual information between the a priori and extrinsic LLRs.
Specifically, plotting the mutual information of the a priori LLRs against the extrinsic 
yields an \emph{EXIT curve}.
It has been shown that matching the variable and check node EXIT curves as closely as possible through selection
of variable node degree yields good LDPC decoding performance \cite{ashikhmin:2004}.

\comment{The optimization developed in this section follows the framework for LDPC code optimization given
described in \cite{brink:2004}.
A soft-output demodulator which produces LLRs for received bits may be modeled jointly
with the LDPC variable nodes to produce an EXIT curve characterizing the demodulator and the variable node decoders.}
LDPC decoding performance may be optimized by matching the EXIT characteristic of the combined demodulator and variable node decoder with the check node decoder.
 In this section we develop optimized LDPC codes having EXIT characteristics matched to the DNC relay demodulator developed in Section \ref{sec:demod}.
\comment{The analytical details of optimization are the same as \cite{brink:2004}, whereas the novelty is in
incorporating the relay demodulator to optimize performance for the system developed in this work. }
The performance of the optimized codes is compared to standard codes and the \comment{achievable rates} calculated in
Section \ref{sec:capacity}.

The variable node degrees for a code are denoted by the set 
$\{d_{v,1}, ..., d_{v,D}\}$, where $d_{v,i}$ is the $i$-$th$ degree and $D$ is the
number of distinct degrees.
The degree distribution is defined as the set of variable node degrees and the 
number of nodes taking a particular degree
$V = \{ d_{v,1}: o_1, ..., d_{v,D}: o_d, d_c\}$
where $o_i$ is the number of variable nodes of degree $d_{v,i}$.


A valid degree distribution satisfies the constraints imposed by the LDPC code parameters.
The total number of edges incident on the variable and check nodes must be the same.
The number of edges incident on variable nodes having degree $d_{v,i}$ is $e_{v,i} = o_i d_{v,i}$,
thus, the total number of edges incident on all variable nodes is
\begin{align}\label{eqn:vnd_edges}
e_{v} = \sum_{i=1}^{D} o_i d_{v,i}.
\end{align}
\noindent 
We consider LDPC codes having a single check node degree, described as
\emph{check regular} codes.
The total number of edges incident on the check nodes is then $e_{c} = d_{c} (N - K)$.
Equating $e_v$ and $e_c$ and rearranging,
\begin{align}\label{eqn:edge_const}
\sum_{i=1}^{D} \frac{o_i d_{v,i}}{d_c (N-K)} = 1.
\end{align}
\noindent The degree distribution of any given parity check matrix must have values 
of $N$, $K$ and $V$ that satisfy (\ref{eqn:edge_const}).
The design challenge is to select degree distributions that optimize
error rate performance for particular channels and relay receiver configurations.

Code optimization is performed as follows.  A range of variable node degree distributions is considered
which satisfy the edge constraint given by Eq. (\ref{eqn:edge_const}).
The EXIT curves for the combined demodulator and variable node decoder
and check node decoder are generated via the Monte Carlo method for all degree distributions.
The check node decoder EXIT curve is completely specified by the check node degree.
Simulation is performed for a range of $\mathcal{E}_b/N_0$ values, noting
the value at which the demodulator and variable node and check node decoder curves intersect.
The highest $\mathcal{E}_b/N_0$ for which the curves intersect is defined as the \emph{EXIT threshold}.
The degree distributions are sorted from lowest EXIT threshold to highest. 
LDPC parity check matrices are realized and simulated starting with the lowest degree distribution
and ending when a code is found that performs better than the standard code.
Note that a more aggressive search may be performed by simulating additional degree distributions.

\begin{table*}[!t]
  \centering
  \caption{LDPC variable node degree optimization results. 
  The SNRs required to reach a BER of $10^{-4}$
  for optimized and standard code simulation are given in columns opt \emph{Opt.} and \emph{Std.} respectively.
  Degree distribution is defined as $V = \{ d_{v,1}:o_1, d_{v,2}:o_2, d_{v,3}:o_3, d_c\}$, where $d_{v,i}$ denotes the $i$-th degree, $o_i$ is the number of nodes taking that degree, and $d_c$ is the check node degree.
  Achievable rates are listed in the columns titled ``Ach. Rate''.}

\vspace{-3mm}
  \begin{tabular}{|l|l|cccc|cccc|}
    \hline
    & & \multicolumn{8}{|c|}{Code Rate ($r$)}\\
    \cline{3-10}
& & \multicolumn{4}{|c|}{$3/5$} & \multicolumn{4}{c|}{$2/5$} \\
    \cline{3-10}
    & & & \multicolumn{2}{c}{Simulated (dB)} & Ach. Rate & & \multicolumn{2}{c}{Simulated (dB)} & Ach. Rate \\
    Channel & M & V & Opt. & Std. & (dB) & V & Opt. & Std. & (dB)\\
    \hline
    AWGN & 2 & 2:6480, 3:7290, 15:2430, 11 & 9.78 & 9.89 & 9.41 & 2:9720, 3:4050, 11:2430, 6 & 9.93 & 9.98  & 9.52 \\
         & 4 & 2:6480, 3:8640, 30:1080, 11 &6.65 & 6.99 & 6.29 & 2:9720, 4:5760, 22:720, 6& 7.09 & 7.17  & 6.57 \\
         & 8 & 2:6480, 3:8991, 43:729, 11& 5.15 & 5.53 & 4.66 & 2:9720, 3:5670, 27:810, 6 & 5.54& 5.86 & 5.07 \\
    \hline
    Rayleigh,& 2 & 2:6480, 4:8640, 22:1080, 11 &17.4  & 17.8 &16.2  & 2:9720, 3:4860, 15:1620, 6 & 14.8& 15.0  & 13.9 \\
    Partial CSI  & 4    & 2:6480, 3:8262, 23:1458, 11 &13.0  & 13.8 &11.7  & 2:9720, 4:5760, 22:720, 6 & 11.0  & 11.3& 9.94 \\
         & 8    & 2:6480, 3:8640, 30:1080, 11 &10.5  & 11.6 &9.40  & 2:9720, 3:4860, 15:1620, 6 &9.18  & 9.44& 8.02 \\
    \hline
    Rayleigh, & 2 & 2:6480, 3:7290, 15:2430, 11 &17.4 & 17.9 & 16.4 & 2:9720, 3:4536, 13:1944, 6& 15.1& 15.3 & 14.3 \\
    No CSI     & 4 & 2:6480, 3:8262, 23:1458, 11 &13.1 & 13.9 & 12.1 & 2:9720, 4:5832, 24:648, 6& 11.4& 11.6& 10.4 \\
         & 8 & 2:6480, 3:8640, 30:1080, 11 & 10.9 & 11.8 & 9.83  & 2:9720, 4:5832, 24:648, 6& 9.55& 9.85& 8.46 \\
    \hline
  \end{tabular}
  \label{tbl:exit_opt_dnc}
  \vspace{-4mm}
\end{table*}

\vspace{-4mm}
\subsection{Optimization Results}
\vspace{-0mm}

This subsection presents the results of EXIT-based LDPC code optimization.
The performance of the optimized codes is compared against standard codes.
Optimized variable node degree distributions are used to realize parity check matrices.
Error rate performance for the optimized codes is computed via Monte Carlo simulation.

Optimization is performed for several cases of receiver configuration,
channel state information, and code rate. 
Specifically, modulation orders $M=\{2,4,8\}$, Rayleigh fading with and without CSI, AWGN,
and code rates $r = \{ 3/5, 2/5 \}$ are considered.
The code length and check node degrees are chosen the same as DVB-S2 to facilitate comparison.
The code length is $N=16200$.
All codes are check-regular.
At code rate $3/5$, the check node degree is $d_c = 11$, and at rate $2/5$ $d_c = 6$.

All codes satisfy the \emph{extended irregular repeat-accumulate} (eIRA) constraint, simplifying
encoding and decoding complexity \cite{yang:2004}.
The eIRA constraint is implemented by partitioning the parity check matrix as
$ \mathbf{H} = [ \mathbf{H}_1 | \mathbf{H}_2 ]$,
where $\mathbf{H}_2$ has dual-diagonal eIRA structure
with $(N-K)$ rows and $(N-K)$ columns.
To preserve the complexity benefits of eIRA, we retain $\mathbf{H}_2$ in the optimized codes 
and consider optimizing the variable node degrees corresponding to $\mathbf{H}_1$.
Retaining $\mathbf{H}_2$ places constraints on $V$ such that optimized codes based on the DVB-S2
have $d_{v,1} = 2$.
All other degrees may be chosen freely.


A range of degree distributions is considered for each code and receiver configuration, and 
the best performing degree distribution under each configuration is realized and simulated.
The number of distinct variable node degrees is
$D=3$, and the degrees considered are all unique combinations of
 $d_{v,1} = 2$, $d_{v,2} \in \{ 2, 4, ... , 99 \}$, 
and $d_{v,3} \in \{ d_{v,2}+1, d_{v,2}+2, ..., d_{v,2}+98 \}$ which satisfy the constraints
for realizable codes described in subsection \ref{subsec:opt_vnd}.
The EXIT threshold is determined for each degree distribution, and
the degree distributions are sorted by EXIT threshold from lowest to highest threshold value.
Starting with the lowest threshold value, degree distributions are realized as parity check matrices 
and simulated until a code is discovered which performs better than the standard.
The resulting degree distributions are shown in Table \ref{tbl:exit_opt_dnc}.


For each degree distribution an LDPC parity check matrix $\mathbf{H}$ is generated by the following heuristic\footnote{Software for generating parity check matrices is at  \url{http://www.cs.utoronto.ca/~radford/ftp/LDPC-2012-02-11/index.html}}.
The submatrix $\mathbf{H}_1$ having $N-K$ rows and $K$ columns is initialized 
to contain all zeros.
For a particular degree distribution $V$, $\mathbf{H}_1$ will contain $o_2$ columns
having $d_{v,2}$ $1$'s, and $o_3$ columns containing $d_{v,3}$ $1$'s.
The total number of $1$'s in $\mathbf{H}_1$ is then $T = d_{v,2}o_2 + d_{v,3}o_3$.
The pool of $T$ $1$'s are assigned to rows as evenly as possible, with remainders assigned
to rows uniformly at random.
In the case that the column weights cannot be satisfied
by the available pool of $T$ ones, $1$'s are assigned at random
to satisfy the column weights.
Additional $1$'s are added to eliminate rows which have weight zero or one.
The position for additional ones are selected uniformly at random from within
the positions containing zeros.
The resulting matrix $\mathbf{H}_1$ is concatenated with the eIRA matrix $\mathbf{H}_2$
to form the parity check matrix.

The optimization results are shown in Table \ref{tbl:exit_opt_dnc}.
At an operating BER of $10^{-4}$, the optimized codes outperform the standard DVB-S2 codes 
for all receiver configurations and channels.
At rate $r=3/5$, the optimal variable node degrees increase with modulation order, while
at rate $r=2/5$ the degrees are more evenly distributed.
The improvement of the optimized codes over standard is greater at rate $3/5$ than $2/5$.
In all cases, the improvement over standard increases with modulation order.
The performance gap of the optimized codes to the \comment{achievable rate} varies between approximately $0.5-1$ dB.

\begin{figure}[!t]
  \centering
  \vspace{0.225cm}
  \includegraphics[width=\columnwidth]{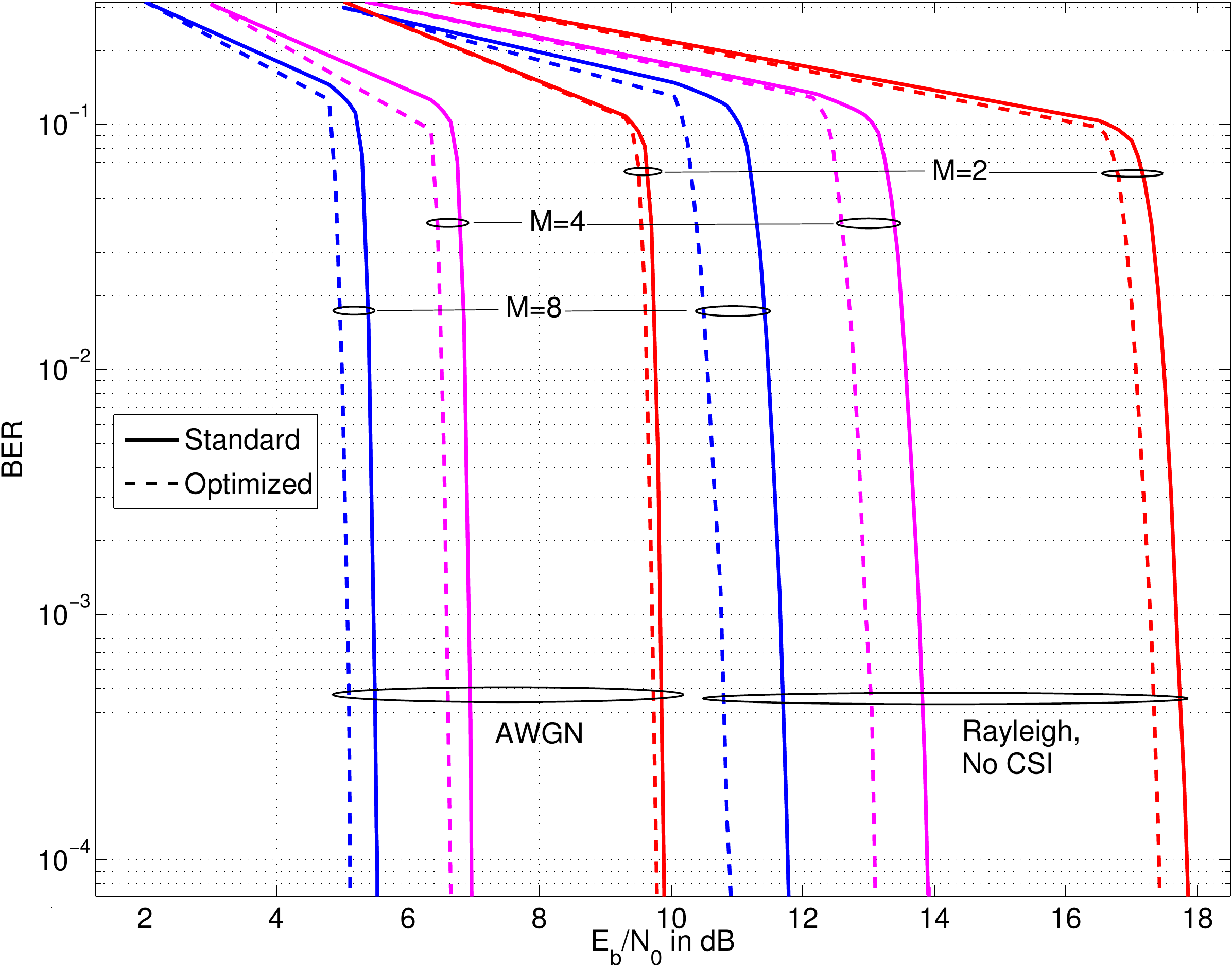}
  \caption{LDPC-coded BER performance at the relay using optimized channel codes for digital network coding.
    The channel code rate is $r=r_M=3/5$.
    Performance is simulated in AWGN and Rayleigh fading with no CSI at the relay.
The frame length is $N=16200$ bits. Modulation orders $M=\{2,4,8\}$ are considered.}
  \label{fig:dnc_opt}
  \vspace{-7mm}
\end{figure}

Simulated bit error rate performance for the optimized codes at rate $r=3/5$ is shown in Fig. \ref{fig:dnc_opt}.
The BER was simulated in AWGN and Rayleigh fading with no CSI at modulation orders $M=\{2,4,8\}$.
One hundred decoding iterations were performed, as a higher number of iterations conferred no additional benefit.
The performance improvement of the optimized codes over the standard codes is nearly constant in the waterfall region.

\vspace{-3mm}


\section{Conclusion}\label{sec:conc}

\emph{Digital network coding} (DNC) is a variant of \emph{physical-layer network coding} where the relay computes the exclusive-or (XOR) of the bits transmitted by the \commentB{terminals}.
\commentB{In this paper we developed a noncoherent modulation and coding system for DNC using multitone FSK.
A novel soft-output demodulator was developed for the relay, and the \emph{achievable exchange rate} was quantified.
The relay receive architecture iterates between the demodulator and LDPC channel decoder to achieve bit-error rate performance that approaches the achievable rate.
DNC was compared against \emph{link-layer network coding} (LNC), a protocol where the \commentB{terminals} transmit to the relay using separate channel resources with no interference. 
The achievable rate analysis revealed that there is a threshold rate above which DNC exhibits better energy efficiency than LNC, and below which LNC efficiency is best.
For DNC, increasing the modulation order from $M=2$ to $M=4$ yields as much as $3$ dB energy efficiency gain, demonstrating the utility of $M$-ary FSK.  Additionally, higher-order FSK exhibits greater energy efficiency gain over binary FSK for noncoherent DNC than for the single-source, single destination point-to-point channel.
A simulation campaign investigated the error rate performance at the relay for DNC and LNC. In particular, several modulation orders were simulated with and without fading amplitude knowledge and using LDPC channel coding.
}
Optimized LDPC codes for DNC were generated by an EXIT curve-fitting process.
Variable node degree distributions were discovered which closely match the EXIT characteristics
of the variable nodes to the check nodes.
The optimized codes outperform well-known standard codes by up to $1.1$ dB, and perform within $0.7$ dB of \comment{achievable rate}.

\balance

\bibliographystyle{IEEEtran}
\bibliography{bibliography}

\begin{thebibliography}{10}
\providecommand{\url}[1]{#1}
\csname url@rmstyle\endcsname
\providecommand{\newblock}{\relax}
\providecommand{\bibinfo}[2]{#2}
\providecommand\BIBentrySTDinterwordspacing{\spaceskip=0pt\relax}
\providecommand\BIBentryALTinterwordstretchfactor{4}
\providecommand\BIBentryALTinterwordspacing{\spaceskip=\fontdimen2\font plus
\BIBentryALTinterwordstretchfactor\fontdimen3\font minus
  \fontdimen4\font\relax}
\providecommand\BIBforeignlanguage[2]{{%
\expandafter\ifx\csname l@#1\endcsname\relax
\typeout{** WARNING: IEEEtran.bst: No hyphenation pattern has been}%
\typeout{** loaded for the language `#1'. Using the pattern for}%
\typeout{** the default language instead.}%
\else
\language=\csname l@#1\endcsname
\fi
#2}}

\bibitem{ahlswede:2000}
R.~Ahlswede, N.~Cai, S.~Li, and R.~Yeung, ``Network information flow,''
  \emph{IEEE Trans. Inform. Theory}, vol.~46, pp. 1204--1216, July 2000.

\bibitem{hausl:2006}
C.~Hausl and J.~Hagenauer, ``Iterative network and channel decoding for the
  two-way relay channel,'' \emph{Proc. IEEE Int. Conf. on Commun.}, vol.~4, pp.
  1568--1573, June 2006.

\bibitem{zhang2:2006}
S.~Zhang, S.~C. Liew, and P.~P. Lam, ``Physical-layer network coding,''
  \emph{Proc. MobiComm}, pp. 358--365, 2006.

\bibitem{2011:torrieri_princ}
D.~Torrieri, \emph{Principles of Spread-Spectrum Communication Systems},
  2nd~ed.\hskip 1em plus 0.5em minus 0.4em\relax Springer Publishing Company,
  Inc., 2011.

\bibitem{wu:2014}
M.~Wu, F.~Ludwig, M.~Woltering, D.~Wuebben, A.~Dekorsy, and S.~Paul, ``Analysis
  and implementation for physical-layer network coding with carrier frequency
  offset,'' in \emph{Int. ITG Workshop on Smart Antennas}, Mar. 2014, pp. 1--8.

\bibitem{sorensen:2009}
J.~S{\o}rensen, R.~Krigslund, P.~Popovski, T.~Akino, and T.~Larsen, ``Physical
  layer network coding for {FSK} systems,'' \emph{IEEE Commun. Lett.}, vol.~13,
  no.~8, pp. 597--599, Aug. 2009.

\bibitem{vtf:2011}
M.~C. Valenti, D.~Torrieri, and T.~Ferrett, ``Noncoherent physical-layer
  network coding with {FSK} modulation: Relay receiver design issues,''
  \emph{IEEE Trans. Commun.}, vol.~29, no.~9, Sept. 2011.

\bibitem{zhang:2014}
D.-Y. Zhang, Q.-Y. Yu, W.-X. Meng, and C.~Li, ``2{FSK} modulation for multiuser
  physical-layer network coding network,'' \emph{Proc. IEEE Int. Conf. on
  Commun.}, pp. 514--519, June 2014.

\bibitem{dang:2016}
X.~Dang, Z.~Liu, B.~Li, and X.~Yu, ``Noncoherent multiple-symbol detector of
  binary {CPFSK} in physical-layer network coding,'' \emph{IEEE Commun. Lett.},
  vol.~20, no.~1, pp. 81--84, Jan. 2016.

\bibitem{yu:2016}
Q.-Y. Yu, D.-Y. Zhang, H.-H. Chen, and W.-X. Meng, ``Physical-layer network
  coding systems with {MFSK} modulation,'' \emph{IEEE Trans. Veh. Technol.},
  vol.~65, no.~1, pp. 204--213, Jan. 2016.

\bibitem{ferrett:2011}
T.~Ferrett, M.~C. Valenti, and D.~Torrieri, ``Noncoherent digital network
  coding using multi-tone {CPFSK} modulation,'' \emph{Proc. IEEE Military
  Commun. Conf.}, pp. 299--304, Nov. 2011.

\bibitem{ferrett:2013}
------, ``An iterative noncoherent relay receiver for the two-way relay
  channel,'' \emph{Proc. IEEE Int. Conf. on Commun.}, pp. 5903--5908, June
  2013.

\bibitem{ferrett:2015}
T.~Ferrett and M.~C. Valenti, ``{LDPC} code design for noncoherent physical
  layer network coding,'' \emph{Proc. IEEE Int. Conf. on Commun.}, pp.
  2054--2059, June 2015.

\bibitem{cui:2009}
T.~Cui, F.~Gao, and C.~Tellambura, ``Differential modulation for two-way
  wireless communications: a perspective of differential network coding at the
  physical layer,'' \emph{IEEE Trans. Commun.}, vol.~57, no.~10, pp.
  2977--2987, Oct. 2009.

\bibitem{zhu:2012}
K.~Zhu and A.~G. Burr, ``A simple non-coherent physical-layer network coding
  for transmissions over two-way relay channels,'' \emph{Proc. IEEE Global
  Commun. Conf.}, pp. 2268--2273, Dec. 2012.

\bibitem{zhang:2008}
S.~Zhang, S.~C. Liew, and L.~Lu, ``Physical layer network coding schemes over
  finite and infinite fields,'' \emph{IEEE Global Telecommun. Conf.}, pp. 1--6,
  Dec. 2008.

\bibitem{caire:1998}
G.~Caire, G.~Taricco, and E.~Biglieri, ``Bit-interleaved coded modulation,''
  \emph{IEEE Trans. Inform. Theory}, vol.~44, no.~3, pp. 927--945, May 1998.

\bibitem{li:1997}
X.~Li and J.~A. Ritcey, ``Bit-interleaved coded modulation with iterative
  decoding,'' \emph{IEEE Commun. Lett.}, vol.~1, no.~6, Nov. 1997.

\bibitem{zhang:2009}
S.~Zhang and S.~C. Liew, ``Channel coding and decoding in a relay system
  operated with physical-layer network coding,'' \emph{IEEE J. Select. Areas
  Commun.}, vol.~27, no.~5, pp. 788--789, June 2009.

\bibitem{li_zhang:2013}
X.~Li, S.~Zhang, and G.~Qian, ``Mapping and coding design for channel coded
  physical-layer network coding,'' \emph{IEEE Int. Workshop on High Mobility
  Wireless Commun.}, pp. 120--125, Nov. 2013.

\bibitem{valenti_xiang:2012}
M.~C. Valenti and X.~Xiang, ``Constellation shaping for bit-interleaved {LDPC}
  coded {APSK},'' \emph{IEEE Trans. Commun.}, vol.~60, no.~10, pp. 2960--2970,
  July 2012.

\bibitem{tanc:2013}
A.~K. Tanc, T.~M. Duman, and C.~Tepedelenlioglu, ``Design of {LDPC} codes for
  two-way relay systems with physical-layer network coding,'' \emph{IEEE
  Commun. Lett.}, vol.~17, no.~12, pp. 2356--2359, Dec. 2013.

\bibitem{huang:2013}
T.~Huang, T.~Yang, J.~Yuan, and I.~Land, ``Design of irregular
  repeat-accumulate coded physical-layer network coding for gaussian two-way
  relay channels,'' \emph{IEEE Trans. Commun.}, vol.~61, no.~3, March 2013.

\bibitem{brink:2004}
S.~ten Brink, G.~Kramer, and A.~Ashikhmin, ``Design of low-density parity-check
  codes for modulation and detection,'' \emph{IEEE Trans. Commun.}, vol.~52,
  no.~4, pp. 670--678, April 2004.

\bibitem{valenti:2005}
M.~C. Valenti and S.~Cheng, ``Iterative demodulation and decoding of turbo
  coded ${M}$-ary noncoherent orthogonal modulation,'' \emph{IEEE J. Select.
  Areas Commun.}, vol.~23, no.~9, pp. 1738--1747, Sept. 2005.

\bibitem{proakis:2008}
J.~G. Proakis and M.~Salehi, \emph{Digital Communications}, 5th~ed.\hskip 1em
  plus 0.5em minus 0.4em\relax New York, NY: McGraw-Hill, Inc., 2008.

\bibitem{dahlman:2011}
E.~Dahlman, S.~Parkvall, and J.~Sk{\"o}ld, \emph{4G LTE/LTE-Advanced for Mobile
  Broadband}.\hskip 1em plus 0.5em minus 0.4em\relax Oxford: Academic Press,
  2011.

\bibitem{benedetto:1998}
S.~Benedetto, G.~Montorsi, D.~Divsalar, and F.~Pollara, ``Soft-input
  soft-output modules for the construction and distributed iterative decoding
  of code networks,'' \emph{Eur. Trans. Telecommun.}, vol.~9, no.~2, pp.
  155--172, Mar.-Apr. 1998.

\bibitem{valenti:2009}
M.~C. Valenti, D.~Torrieri, and T.~Ferrett, ``Noncoherent physical-layer
  network coding using binary {CPFSK} modulation,'' \emph{Proc. IEEE Military
  Commun. Conf.}, pp. 1--7, Oct. 2009.

\bibitem{noori:2012}
M.~Noori and M.~Ardakani, ``On symbol mapping for binary physical-layer network
  coding with {PSK} modulation,'' \emph{IEEE Trans. Wireless Commun.}, vol.~11,
  no.~1, pp. 21--26, Jan. 2012.

\bibitem{zhang:2010}
S.~Zhang, S.~C. Liew, H.~Wang, and X.~Lin, ``Capacity of two-way relay
  channel,'' \emph{4th Int. Conf. AccessNets}, pp. 219--231, Nov. 2010.

\bibitem{cover:2006}
T.~M. Cover and J.~A. Thomas, \emph{Elements of Information Theory}.\hskip 1em
  plus 0.5em minus 0.4em\relax Wiley-Interscience, 2006.

\bibitem{dvbs2:2013}
``Digital video broadcasting {(DVB)},'' \emph{ETSI EN 302 307 V1.3.1}, 2013.

\bibitem{ashikhmin:2004}
A.~Ashikhmin, G.~Kramer, and S.~ten Brink, ``Extrinsic information transfer
  functions: Model and erasure channel properties,'' \emph{IEEE Trans. Inform.
  Theory}, vol.~50, no.~11, Nov. 2004.

\bibitem{yang:2004}
M.~Yang, R.~W. E., and Y.~Li, ``Design of efficiently encodable moderate-length
  high-rate irregular {LDPC} codes,'' \emph{IEEE Trans. Commun.}, vol.~52,
  no.~4, pp. 564--571, April 2004.

\end{thebibliography}

\vspace{-1cm}

\begin{IEEEbiography}
[{\includegraphics[width=1in,height=1.25in,clip,keepaspectratio]{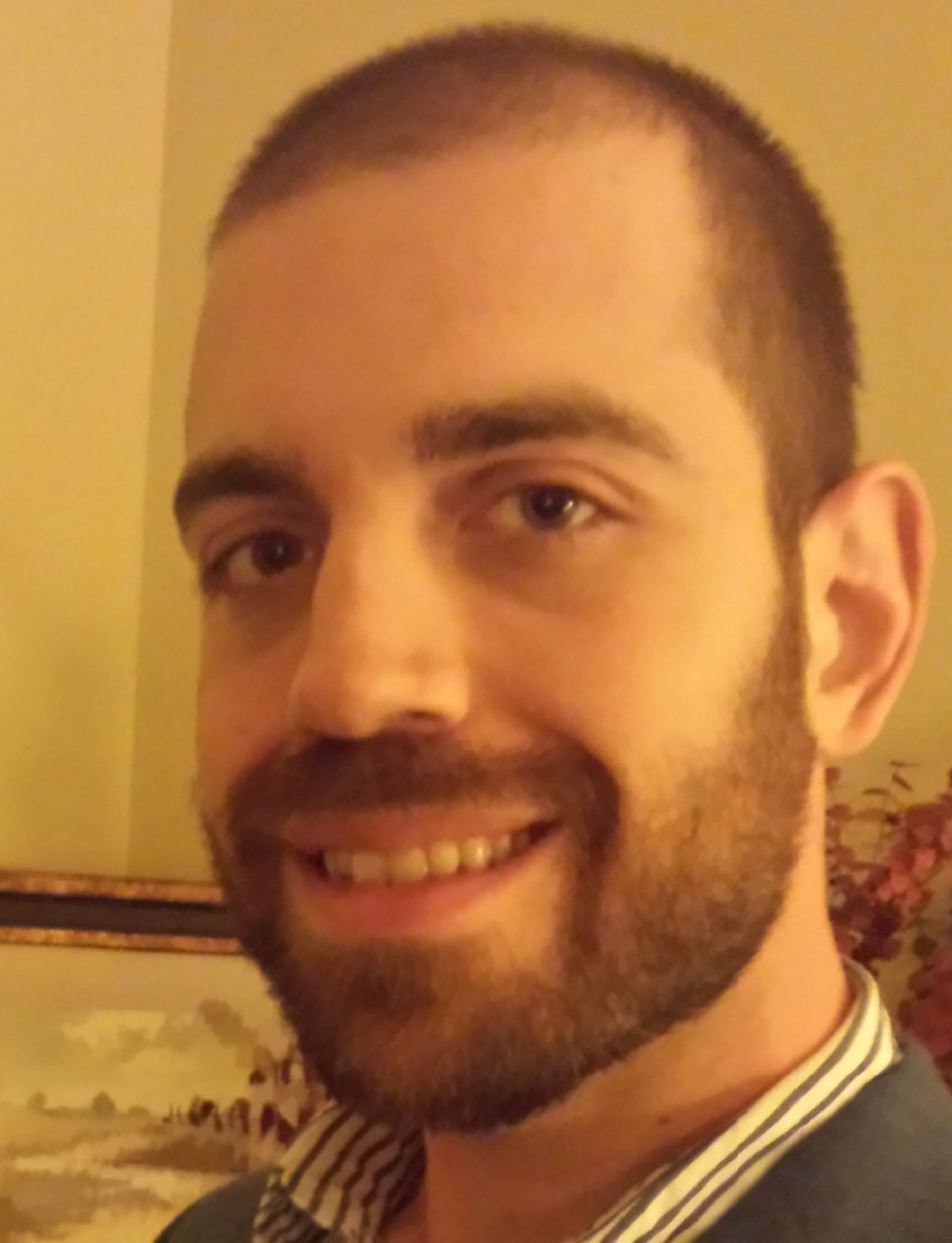}}]{Terry Ferrett}
is a postdoctoral fellow at West Virginia University, Morgantown, WV.  He completed his Ph.D., M.S.E.E., B.S.E.E and B.S.C.E. at West Virginia University.  He is the architect of a cluster computing resource utilized by electrical engineering students at West Virginia University to conduct communication theory research. His research interests are network coding, digital receiver design, the information theory of relay channels, cloud and cluster computing.
\end{IEEEbiography}


\vspace{-1cm}


\begin{IEEEbiography}
[{\includegraphics[width=1in,height=1.25in,clip,keepaspectratio]{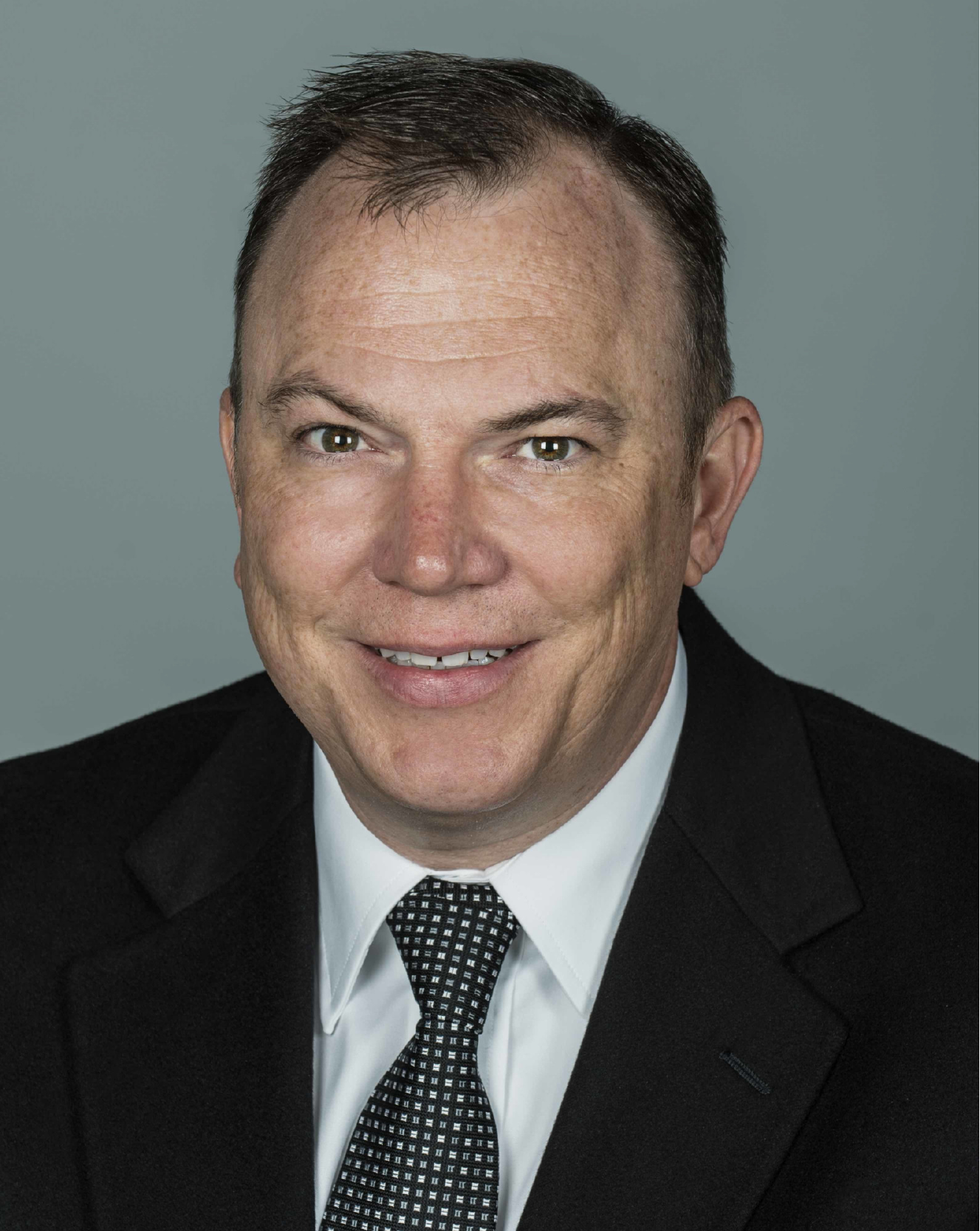}}]{Matthew C. Valenti}
(M'92 - SM'07 - F'18) received the M.S.E.E. degree from the Johns Hopkins University, Baltimore, MD, USA, and B.S.E.E. and Ph.D. degrees from Virginia Tech, Blacksburg, VA, USA. He has been a Faculty Member with West Virginia University since 1999, where he is currently a Professor and the Director of the Center for Identification Technology Research. His research interests are in wireless communications, cloud computing, and biometric identification.  He is active in the organization and oversight of several ComSoc sponsored IEEE conferences, including MILCOM, ICC, and Globecom. He was Chair of the ComSoc Communication Theory Technical committee from 2015-2016, was TPC chair for MILCOM'17, is Chair of the Globecom/ICC Technical Content (GITC) Committee (2018-2019), and is TPC co-chair for ICC'21 (Montreal).  He was an Electronics Engineer with the U.S. Naval Research Laboratory, Washington, DC, USA. Dr. Valenti is registered as a Professional Engineer in the state of West Virginia.
\end{IEEEbiography}

\end{document}